\documentclass{aa}
\usepackage{graphicx}
\usepackage{natbib}
\usepackage{amsmath}
\usepackage{amssymb}
\def\void#1{{}} 
\def\arcmin{$\, ^{\prime}$}
\def\arcsec{$\, ^{\prime\prime}$}

\begin{document}
\title{ESO Imaging survey: Optical Deep Public Survey \thanks{Based on
observations carried out at the European Southern Observatory, La
Silla, Chile under program Nos. 164.O-0561, 169.A-0725, and
267.A-5729.}}

\author{A.~Mignano\inst{1,2,3} \and J.-M.~Miralles\inst{2,4} \and L. da
Costa\inst{2,5} \and L. F. Olsen\inst{6,2,7} \and I. Prandoni\inst{3} \and S. Arnouts\inst{8} \and
C. Benoist\inst{6}  \and R. Madejsky\inst{9} \and
\and R. Slijkhuis\inst{2} 
\and S. Zaggia\inst{10} }

\offprints{A. Mignano, \email{ amignano@ira.inaf.it}}
\institute{Dipartimento di Astronomia, Universita' di Bologna, via Ranzani~1, I-40126
Bologna, Italy
\and
European Southern Observatory, Karl-Schwarzschild-Str.~2, 85748
Garching b. M\"unchen, Germany
\and INAF - Istituto di Radioastronomia, Via Gobetti~101, I-40129 Bologna, 
Italy
\and T\`ecniques d'Avantguarda, Avda. Carlemany 75, AD-700 Les Escaldes, Andorra
\and 
Observatorio Nacional, Rua Gel. Jose Cristino 77, Rio de Janeiro, R.J.,Brazil
\and
Observatoire de la C\^ote d'Azur, Laboratoire Cassiop\'ee, BP4229, 06304 Nice Cedex 4, France
\and Dark Cosmology Centre, Niels Bohr Institute, University of Copenhagen, Juliane Maries Vej 30, DK-2100 Copenhagen, 
      Denmark
\and
Laboratoire d'astrophysique de Marseille, Traverse du Siphon, BP 8, 13376 Marsei
lle Cedex 12, France
\and
Universidade Estadual de Feira de Santana, Campus Universit\'{a}rio, Feira de Santana, BA, Brazil
\and
Osservatorio
Astronomico di Trieste, Via G.B. Tiepolo, 11, I-34131 Trieste, Italy}

\date{Received -; Accepted -}

\abstract{This paper presents new five passbands ($UBVRI$) optical
  wide-field imaging data accumulated as part of the DEEP Public
  Survey (DPS) carried out as a public survey by the ESO Imaging
  Survey (EIS) project. Out of the 3~square degrees originally
  proposed, the survey covers {\tt 2.75} square degrees, in at least
  one band ( normally $R$), and 1.00 square degrees in five
  passbands. The median seeing, as measured in the final stacked
  images, is $0\farcs97$, ranging from $0\farcs75$ {\tt to}
  $2\farcs0$. The median limiting magnitudes (AB system, 2\arcsec
  aperture, $5\sigma$ detection limit) are $U_{AB}=25.65$,
  $B_{AB}=$25.54, $V_{AB}=$25.18, $R_{AB} =$ 24.8 and $I_{AB}
  =$24.12~mag, consistent with those proposed in the original survey
  design. The paper describes the observations and data reduction
  using the EIS Data Reduction System and its associated EIS/MVM
  library. The quality of the individual images were inspected,
  bad images discarded and the remaining used to produce final image
  stacks in each passband, from which sources have been
  extracted. Finally, the scientific quality of these final images and
  associated catalogs was assessed qualitatively by visual
  inspection and quantitatively by comparison of statistical measures
  derived from these data with those of other authors as well as model
  predictions, and from direct comparison with the results obtained
  from the reduction of the same dataset using an independent (hands-on)
  software system. Finally to illustrate one application of this
survey, the results of a preliminary effort to
  identify sub-mJy radio sources are reported. To the limiting
  magnitude reached in the $R$ and $I$ passbands the success rate
  ranges from 66 to 81\% (depending on the fields).  These data are
  publicly available at CDS\thanks{ Available at
CDS via anonymous ftp to cdsarc.u-strasbg.fr (130.79.128.5) or via
http://cdsweb.u-strasbg.fr/cgi-bin/qcat?J/A+A/}.

} 
\maketitle

\section{Introduction}
\label{sec:introduction}

Deep multi-wavelength observations are a pre-condition for the
selection of suitable galaxy samples for spectroscopic follow-up
observation using the multiplex capability of spectrographs mounted on
large-aperture telescopes. Foreseeing the need for such samples for
the scientific exploitation of multi-object spectrographs such as FORS
and VIMOS, ESO's Survey Working Group (SWG)
recommended the ESO Imaging Survey (EIS, \citet{RenzinidaCosta97})
project to undertake a series of deep, optical/infrared
observations. Based on ideas submitted by the ESO community and
evaluated by the SWG, the DEEP Public Survey (DPS) was designed to
provide a data set that could be used to study the large-scale
distribution of galaxies at $z\sim1$ and to identify a large number of
Ly-$\alpha$ galaxies and,  combined with near-infrared data,
extremely red objects (EROs) for follow-up work using VLT.  In its
entirety this survey comprises three optical and infrared
strategies. The optical part, the focus of the present paper, consists
of a deep survey covering three regions of 1~square degree on the sky,
in the $U$-, $B$-, $V$-, $R$- and $I$ passbands, using the wide-field
imager (WFI) mounted on the ESO/MPG 2.2m telescope at La Silla. Each
of the three DEEP regions --~1,~2,~3~-- is covered by four WFI
pointings (in five passbands) --~a, b, c, d. The regions were selected
both to enable observations year-round and because they overlapped
with regions of other scientific interest. For instance, DEEP1 was
chosen to complement  the sub-mJy ATESP radio survey \citep{Prandoni2000a,Prandoni2000b,Prandoni2005}
carried out
with the Australia Telescope Compact Array (ATCA) covering the region
surveyed by the ESO Slice Project \citep{Vettolani1997}, while DEEP2 included the CDF-S
field. Finally, DEEP3, was chosen in the northern galactic hemisphere,
thus providing an almost year-round coverage\footnote{The text for the
original proposal can be found at
http://www.eso.org/science/eis/documents.html}. The location and the
characteristics of the surveyed regions as well as the planned
limiting magnitudes in each passband can be found in the DPS strategy
page\footnote{http://www.eso.org/science/eis/surveys/strategy\_DPS.html}
available in the EIS web-pages.

This paper is part of a series describing the data releases made
by the EIS project. In \cite{Dietrich2005} (hereafter Paper~I) the
data reduction system and the general procedures adopted are
described. Here the observations, reduction, and science verification
of the optical data taken as part of the DPS,
publicly released late 2004, are reviewed. Furthermore, in this
paper changes to the dataset that have occurred since the original
release, stemming from additional evaluation of the data, are also
described.  Complementing the optical results reported in the present
paper, the results obtained from infrared observations covering these
fields are presented in \cite{Olsen2006}.
Sect.~\ref{sec:observations} describes optical observations while the
reduction and calibration of the data are presented in
Sect.~\ref{sec:reduction}. Final survey products such as stacked
images and science-grade catalogs extracted from them are presented in
Sect.~\ref{sec:products}. The quality of these products is evaluated
in Sect.~\ref{sec:discussion} by comparing statistical measures
obtained from these data with those of other authors.
In this section the results of a preliminary assessment of the
optical identification of sub-mJy radio sources from the ATESP 1.4 GHz
radio survey are also discussed. Finally, in Sect.~\ref{sec:summary}
a brief summary of the paper is presented.

\section{Observations}
\label{sec:observations}

The optical observations were carried out using WFI at the
ESO/MPG-2.2m telescope. WFI is a focal reducer-type mosaic camera
mounted at the Cassegrain focus of the telescope. The mosaic consists
of $4 \times 2$ CCD chips with $2048 \times 4096$ pixels with a
projected pixel size of $0\farcs{238}$, giving a FOV of $8\farcm{12}
\times 16\farcm{25}$ for each individual chip. The chips are separated
by gaps of $23\farcs{8}$ and $14\farcs{3}$ along the right ascension
and declination direction, respectively. The full FOV of WFI is thus
34\arcmin $\times$ 33\arcmin with a filling factor of $95.9$\%.

The data presented in this paper were obtained as part of the ESO
Large Programme: 164.O-0561, carried out in visitor mode and its
service mode continuation (30 hours) 169.A-0725 (Principal
Investigator: J. Krautter, as chair of the SWG) as well as
the service mode program 267.A-5729.

Observations, nearly all in visitor mode, were performed in the $U$-,
$B$-, $V$-, $R$-, and $I$-passbands.  These were split into OBs
consisting of a sequence of five (ten in the $I$-band) dithered
sub-exposures with the typical exposure time given in
Table~\ref{tab:ob}.  Filter
curves can be found in \cite{Arnouts2001} and on the web page
of the La Silla Science Operations
Team\footnote{http://www.ls.eso.org/lasilla/sciops/2p2/E2p2M/WFI/filters}.
Note that in an attempt to improve the performance of the survey, new
$U$ and $I$ filters were purchased and used during the course of the
observations. In addition, there were also changes in the observing
strategy in the $U$-, to conform to the one hour limit for a single OB
imposed by the telescope team, and the $I$-band, to improve the fringing
correction. It is important to point out that the changes in strategy
were unrelated to the change of the actual filter used.

\begin{table}
\center
\caption{Typical OBs used in the observations for the DPS optical survey. The table gives: in Col.~1 the passband; in
Col.~2 the filter id adopting the unique naming convention of the La
Silla Science Operations Team; in Col.~3 the number of exposures in
the OB; in Col.~4 the integration time of the individual sub-exposures
in the OB; and in Col.~5 the total exposure time in seconds.} 
\label{tab:ob}
\begin{tabular}{lcccc}
\hline\hline
Passband &    Filter & \# exposures & $T_{exp}$ & $T_{tot}$ \\
         &   &            & (sec)  & (sec) \\                        
\hline
$U$        &  \#877  &  5  &  900  & 4500 \\
$U$        &  \#841  &  5  &  550  & 2750 \\
$B$        &  \#842  &  5  &  300  & 1500 \\
$V$        &  \#843  &  5  &  300  & 1500 \\
$R$        &  \#844  &  5  &  300  & 1500 \\
$I$        &  \#879  &  5  &  600  & 3000 \\
$I$        &  \#845  & 10  &  300  & 3000 \\
\hline
\end{tabular}
\end{table}

The data from the above programs were accumulated in observations
covering six semesters (ESO periods 64-69) from November 4, 1999
through September 28, 2002, corresponding to 1905 science exposures,
and 246 hours of on-target integration, during 76 nights. \void{About 132
exposures (28 science OBs) in the $V$- and $R$-band were taken in 10
nights from the contributing program. About 34\% of the $V$-band and
26\% of the $R$-band data are from the contributing program.}  With the
data presented in this paper 90\% of the data accumulated by the DPS
program have been reduced and publicly released. The remaining 196
exposures include tests, wrong pointings, as well as discarded
images. About 10\% of the images in all filters have been discarded,
with the exception of the $B$-band.  The data covers 2.75 squares
degrees, in at least one band, and 1.00 square degrees in five
passbands.

Standard star observations are available for 75 out of the 76
nights with science observations. These consist of a total of roughly
500 standard star OBs, nearly all consisting of two images per OB.
While this amounts to slightly more than 10 hours on-target (or 4\% of
the science observations in terms of time), it exceeds the volume of
science frames.  For three nights either no standard fields were observed
(September 27, 2002) or they are not available for a particular filter
(November 29, 2000, $U$-band and June 7, 2002, $I$-band). Depending on
the night, the number of measurements can range from a few to over
300, covering from 1 to 3 Landolt fields \citep{Landolt1992}.  

\section {Data reduction}
\label{sec:reduction}

The WFI data were reduced and calibrated using the EIS data reduction
system and its associated EIS/MVM image processing library version
1.0.1. The latter is designed to automatically process images from
single/multi-chip optical/infrared cameras. In addition to the
standard bias-subtraction, flatfield correction and trimming, the
package includes sophisticated algorithms for background estimation,
de-fringing, astrometric calibration, minimization of chip-to-chip
variations in sensitivity and detection/masking of satellite tracks so
common in wide-field imagers. This software package is publicly
available and can be retrieved from the EIS release
page\footnote{http://www.eso.org/science/eis/survey\_release.html}. Details
about the EIS/MVM library can be found in \cite{Vandame2004}. In
Paper~I the reduction and general handling of WFI surveys through the
EIS data reduction system is described.  In the following the
points specific for the DPS WFI data are described.

The astrometric calibration was derived using the GSC2.2 reference
catalog and a distortion model described by a second order
polynomial. Comparisons with independent astrometric
catalogs yield a typical scatter of 0.2~arcsec \citep{Vandame2004}. 
However, comparisons made using exposures of a globular
cluster, shifted by half a field, strongly suggest that the internal
accuracy is about 70~mas. The astrometric calibration is carried
out on a chip by chip basis, which occasionally may lead to less than
optimal PSF as discussed in Paper~I, which also describes possible
improvements to be implemented to the code in the future. The reduced
images are warped onto a user-defined reference grid (pixel,
projection and orientation), using a third-order Lanczos kernel. The
WCS in the image header is reported using the CD-matrix notation.

The EIS Data Reduction System also includes a photometric pipeline for
automatic extraction of the photometric solutions used to calibrate
reduced images as described in detail in Paper~I.  As mentioned above,
standard stars were observed in 75  nights corresponding to about 500
OBs. These were used to obtain photometric solutions for each night
for calibrating the reduced images.

The EIS reduction system derives a number of photometric solutions for
each night. Depending on the airmass and color coverage between one and
three parameters are fit. Among these solutions the one with the smallest
scatter is considered the best fit and refered to as the ``best solution''.
Table~\ref{tab:phot} summarizes the number of nights with standard
star observations and type of ``best solution'' obtained by the
photometric pipeline for each passband. It can be seen that for all filters,
except $U$\#841, 2- and 3-parameter fits are available, in
principle, allowing for reliable photometric calibration.  In the
case of the $U$\#841 the airmass coverage was insufficient and only
one-parameter fits are available. Neither does the filter have a solution
from the Telescope Team thus hampering further confirmation.

\tabcolsep 0.15cm
\begin{table}
\center
\caption {Number of nights with different types of photometric solution. The table gives in Col.~1 the passband;
in Col.~2 the filter identification; in Col.~3 the number of default
solutions (corresponding to the number of non-photometric nights and/or
nights without standard star observations); in Cols.~4--6 the number
of nights with 1-, 2- or 3-parameter fits, respectively; and in Col.~7
the total number of nights in which standard star observations were
carried out with a given filter.}
\label {tab:phot}
\begin{tabular}{ccccccc}
\hline\hline
Passband & Filter id & default & 1-par & 2-par & 3-par & total \\
\hline
$U$&\#877&3&17&12&6&38 \\
$U$&\#841&0&9&0&0&9 \\
$B$&\#842&0&13&2&2&17 \\
$V$&\#843&0&10&7&6&23 \\
$R$&\#844&5&6&6&2&19 \\
$I$&\#879&1&0&2&1&4 \\
$I$&\#845&6&9&4&7&26 \\
\hline
\end{tabular}
\end{table}

Table~\ref{tab:field-calib} gives the type of fit for the "best solution" obtained for each
field and filter indicating the quality of the photometric calibration. It
can be seen that for all field/filter combinations there is at least one
reduced image calibrated with at least a 1-parameter fit.
In addition, with
exception of the $B$-band data for DEEP1 and DEEP2,\void{ and the
$R$-band data for DEEP3,}  all others have at least a 2-parameter fit
solution, thereby ensuring a reasonable photometric calibration.  More
importantly, the table also shows that in nearly all regions a 3-parameter fit exists
for at least one field thus allowing the
photometric calibration to be bootstrapped or cross-checked to other
fields.

\tabcolsep 0.1cm
\begin{table}
\center
\caption {Type of photometric solution per passband for each observed
  DPS field. The table gives: in Col.~1 the
region; in Col.~2 the field; in Cols.~3--6 the type of best solution
available for a given passband.}
\label{tab:field-calib}
\begin{tabular}{llcccc}
\hline\hline
Region &  Field Name & default & 1-par & 2-par & 3-par \\
\hline
DEEP1&DEEP1a&  &  $U$    $B$  &  $I$  &  $U$    $R$   \\
DEEP1&DEEP1b&  &  $B$  &  $U$  &  $V$    $R$    $I$   \\
DEEP1&DEEP1c&  &  &  $V$    $R$  &   \\
DEEP2&DEEP2a&  &  &  $R$  &   \\
DEEP2&DEEP2b&  &  $B$  &  $U$  &  $V$    $R$    $I$   \\
DEEP2&DEEP2c&  &  $U$    $I$  &  $U$  &  $V$   \\
DEEP2&DEEP2d&  &  $R$  &  &   \\
DEEP3&DEEP3a&  &  $U$    $U$    $R$    $I$  &  &  $B$    $V$   \\
DEEP3&DEEP3b&  &  &  $R$  &  $U$    $B$    $V$    $I$   \\
DEEP3&DEEP3c&  &  &  $B$    $V$    $R$  &  $U$    $I$   \\
DEEP3&DEEP3d&  &  $B$  &  $V$  &  $I$   \\
\hline
\end{tabular}
\end{table}

As part of the photometric calibration, nights with good airmass and
color coverage were used to compute complete photometric solutions,
namely those for which all parameters (zeropoint, extinction and color
term) can be estimated. The median values derived for
these 3-parameter fits are shown in Table~\ref{tab:3par}.  
As mentioned earlier, no such
solution is available for the $U$-band filter \#841.

\begin{table}
\center
\caption{Median values for all photometric solutions based on
  3-parameter fits. The table
lists: in Col.~1 the passband/filter; in Cols.~2--4 the median
zeropoint in Vega System (Zp), extinction, (k), and color term (color) values for all
3-parameter fit solutions.}
 \label{tab:3par}
\begin{tabular}{cccc}
\hline\hline
Passband & Zp &  k &  color  \\
\hline
$U$\#877&22.06&0.5&0.04 \\
$B$\#842&24.58&0.22&0.24 \\
$V$\#843&24.23&0.19&-0.04 \\
$R$\#844&24.49&0.08&-0.01 \\
$I$\#879&23.35&0.02&0.03 \\
$I$\#845&23.18&0.09&0.24 \\
\hline
\end{tabular}
\end{table}

The photometric solutions computed automatically by the EIS data
reduction system were compared with the ``best solution''
recently obtained by the 2p2 Telescope Team. The results of this
comparison are presented in Table~\ref{tab:zcomp} for the general case
of 3-parameter fits.  The solutions are remarkably similar,
despite the fact that they are over two years apart. It is worth
emphasizing that, for the present release, the periods of observations
of standard stars available to the two teams do not coincide. Also
note that there are three (in {\tt $U\#841$, $B\#842$\footnote{The solution for
the $B$-band reported in the web-page does not correspond the same
filter as used here.}  and $I\#845$}) filters EIS has used for which no
solutions have been reported by the telescope team. These cases are not
shown in the table.

\begin{table}
\center
\caption{Comparison between EIS and Telescope Team solutions. The table lists: in Col.~1 the passband; in
Cols.~2-4 the offsets (EIS-Telescope Team) of the zeropoint,
extinction and color term.}
\label{tab:zcomp}
\begin{tabular}{cccc}
\hline\hline
Passband & Zp & k & color  \\
\hline
$U$\#877&0.0&0.02&-0.01 \\
$V$\#843&0.08&0.08&0.09 \\
$R$\#844&0.02&0.01&-0.01 \\
$I$\#879&-0.02&0.02&0.0 \\
\hline
\end{tabular}
\end{table}

The quality of the photometric calibration of the present dataset can
further be assessed from Table~\ref{tab:zpquality}. Considering only 3-parameter fit solutions and
after rejecting 3 sigma outliers one finds that the scatter of the
zeropoints is less than about 0.1 mag, which is a 
reasonable estimate for the accuracy of the absolute photometric calibration of
the DPS survey data. Not surprisingly, larger offsets are found when
2- and 1-parameter fits are included, depending on the passband and
estimator used for extinction and color term.  One sees that
considerable work remains to be done to significantly reduce the
photometric calibration error which is typically of the order of 0.1
mag. A notable exception is the wider $U$\#877, for which the rms
exceeds 0.2 mag, indicating that it will be difficult to obtain a
proper calibration for this filter.

\begin{table}
\center
\caption{Average parameters for all photometric solutions and
constraining the sample to the 3-parameter fits. The table gives:
in Col.~1 the passband/filter combination; in Col.~2 the median
zeropoint in Vega System (Zp) including solutions obtained from fits with an
arbitrary number of free parameters; in Col.~3 the scatter computed
from the distribution of zeropoints (rms); in Col.~4 the largest
offset of a night calibration relative to the median reported,
$(\Delta Zp)_{max}$; in Cols.~5--7 the same as in the three previous
columns, except that now the estimate of the median is computed using
only 3-parameter fits.}
\label{tab:zpquality}
\begin{tabular}{ccccccc}
\hline\hline
Passband &  Zp & rms & $(\Delta Zp)_{max}$ & Zp & rms & $(\Delta Zp)_{max}$  \\
\hline
U\#877&22.02&0.27&-0.72&22.06&0.22&-0.68 \\
U\#841&21.58&0.08&0.11&--&--& -- \\
B\#842&24.64&0.08&-0.21&24.58&0.05&-0.27 \\
V\#843&24.18&0.08&0.16&24.23&0.1&0.21 \\
R\#844&24.52&0.08&-0.17&24.49&0.09&-0.2 \\
I\#879&23.34&0.02&0.02&23.35&0.0&0.03 \\
I\#845&23.14&0.11&-0.3&23.18&0.13&-0.26 \\
\hline
\end{tabular}
\end{table}

As final points regarding the photometric calibration of the DPS data,
it is important to emphasize that the CCD-to-CCD gain variations have
been corrected using the median background values sampled in
sub-regions bordering adjacent CCDs but, as in Paper~I, no
illumination correction has been applied which may lead to relative
magnitude offsets from the center to the borders of the image. \void{
In this respect, it is important to recall that wide-field multi-chip
cameras are complex systems and to obtain a proper calibration depends
upon a full understanding of the different effects. In fact, a number
of improvements to the photometric pipeline still need to be
considered such as: 1) better estimates for the extinction and color
term in the case that a three-parameter fit is not possible; 2) an
investigation of possible residual chip-to-chip gain differences even
though these have been corrected for by the pipeline; 3) possible
effect of the illumination correction; 4) a more complete test of the
automatic extract/match/measure and fit procedures adopted and the
associated configuration parameters (including SExtractor 
\citep{Bertin1996}); 5) the
possible contribution of secondary standards not yet used and which
may provide a more uniform coverage across the detector. A detailed
discussion of these points is beyond the scope of the present paper.
It is also known that large-scale variations due to non-uniform
illumination over the field of view of wide-field instruments exist.
The significance of this effect is passband-dependent and becomes more
pronounced with increasing distance from the optical axis
\citep{Manfroid2001,Koch2004,Vandame2004}. Automated software to
correct for this effect has been developed but due to time constraints
it has not yet been applied to these data.}

The EIS/MVM library was used to convert the available
1905 raw WFI science exposures (171 Gb) into 331 fully calibrated
reduced images, of which 304 were released and are discussed in the
present paper. Of the remaining 27, one was observed with the
narrow-band $I$ filter \# 854 and 26 (1 $U$, 6 $B$, 6 $V$, 1 $R$ and 13 $I$)
images were rejected during visual inspection (grading).  Therefore,
the present paper considers 304 fully calibrated, reduced WFI images
in the following passbands:  1) 118 $U$-band images -- 21 using filter
\#841, with a mean wavelength of 3676 \AA\ (FWHM = 325 \AA), and 97
using the wider filter \#877, central wavelength of 3404 \AA\ (FWHM =
732 \AA); 
2) 38 $B$-band images with a mean wavelength of 4562 \AA\ (FWHM = 990 \AA); 
3) 35 $V$-band images with a mean wavelength of 5395 \AA\ (FWHM = 893 \AA); 
4) 48 $R$-band images with a mean wavelength of 6517 \AA\ (FWHM = 1621 \AA); 
and 5) 65 $I$-band images -- 8 obtained using the $I$ filter
\#879, with a mean wavelength of 8269 \AA\ (5 nights in the period
from March 10, 2002 to June 7, 2002), and 57 using the filter \#845
with a central wavelength of 8643 \AA, and FWHM = 1387 \AA\ (25 nights
in the period from November 4, 1999 to August 21, 2001).  The
available data covers 11 fields of the 12 requested. The
completeness of the observations is listed in
Table~\ref{tab:attributes} for each stacked image and the
corresponding ``completeness'' plots are accessible from the EIS
web-pages. 
The zero point of each of the reduced images was computed based on the
photometric solution of the night of observations. If no solution
could be obtained, either because of missing data or due to a
non-photometric night, a default zero point was assigned as detailed in
Paper I.

Before being released all the science images were examined
by eye and graded from
A (best) to D (worst).  Out of 331 reduced images covering the
selected DPS fields, 142 (43\%) were graded A, 121 (37\%) $B$, 41
(12\%) C and 26 (8\%) D. The grade distribution is a function of the
passband as can be seen from Table~\ref {tab:grade-distribution}
showing the grade breakdown for each filter. The table shows
that for $BVR$ bands over 75\% of the products are graded A, while for
$U$-band the typical grade is $B$ and in $I$-band all grades are
roughly equally represented.  In the $U$-band images (both filters),
the low grades are mainly due to clearly visible electronic noise
because of the low level counts of the background in most exposures
(550-900 seconds).  In the $I$-band the low grades are usually caused
by residual fringing. The latter is especially true in the case of
filter \#845 due to the long integration time used per exposure,
leading to inadequate estimate of the fringing map. Specific details about the fringing removal
procedure of EIS/MVM can be found in \cite{Vandame2004}.
Even though some
reduced images have grades C they are still useful in building the
final stack, therefore only images with grade D are rejected.  
These C graded
images may present a higher photometric noise but the
estimated variations in flux are less than 10\%.
A graphic display of the grade distribution can be accessed from the
release page of this survey in the EIS web-pages.  Note that images
with grade D were not released, since they have no scientific value,
and were discarded from the stacking process discussed below.

It is important to emphasize that the grade attributed to a product
depends on the stage of the process. In fact, the grade of the
reduced images does not reflect the quality of the final ``stacked''
image, which in general have considerably better grades
(see below).  For instance, the reduced images still show a number
of cosmic ray hits, because the construction of reduction blocks (RB)
is optimized for removing cosmic ray features in the final stacks. In
most cases, this implies that each field and filter is observed with
at least 3 RBs. Therefore, in most cases the stack blocks (SB) consist
of at least 3 input frames, allowing for the use of a thresholding
procedure to remove cosmic ray hits from the final stacked image.  In
contrast to Paper~I, over 60\% of the reduction blocks consist of 5 or
less frames. This has the effect of leaving the imprint of the
inter-chip gaps, leading to a larger number of cosmic rays in the
reduced images and in the final stack at the location of the
inter-chip gaps (see Paper~I). As in the case of Paper~I the masking
of satellite tracks seemed to have worked remarkably well.

\begin{table}
\center
\caption {Distribution of grades per passband. The table gives: in
Col.~1 the passband; in Col.~2 the ESO filter number; in Col.~3 the
number of images; in Cols.~4--7 the ratio of images in a given grade to
the total number of images taken with that filter.}

\label {tab:grade-distribution}
\begin{tabular}{ccccccc}
\hline\hline
Passband & Filter & \# reduced  & A & B & C & D \\
\hline
$U$&\#877&97&0.22&0.69&0.09&0.0\\
$U$&\#841&22&0.05&0.91&0.00&0.05\\
$B$&\#842&44&0.82&0.02&0.02&0.14\\
$V$&\#843&41&0.73&0.02&0.1&0.15\\
$R$&\#844&48&0.77&0.19&0.04&0.0\\
$I$&\#879&8&0.63&0.25&0.13&0.0\\
$I$&\#845&70&0.17&0.3&0.34&0.19\\
\hline
\end{tabular}
\end{table}

\section {Final products}
\label{sec:products}

\subsection {Images}

\begin{figure*}[h]
  \center
\resizebox{\hsize}{!}{\includegraphics{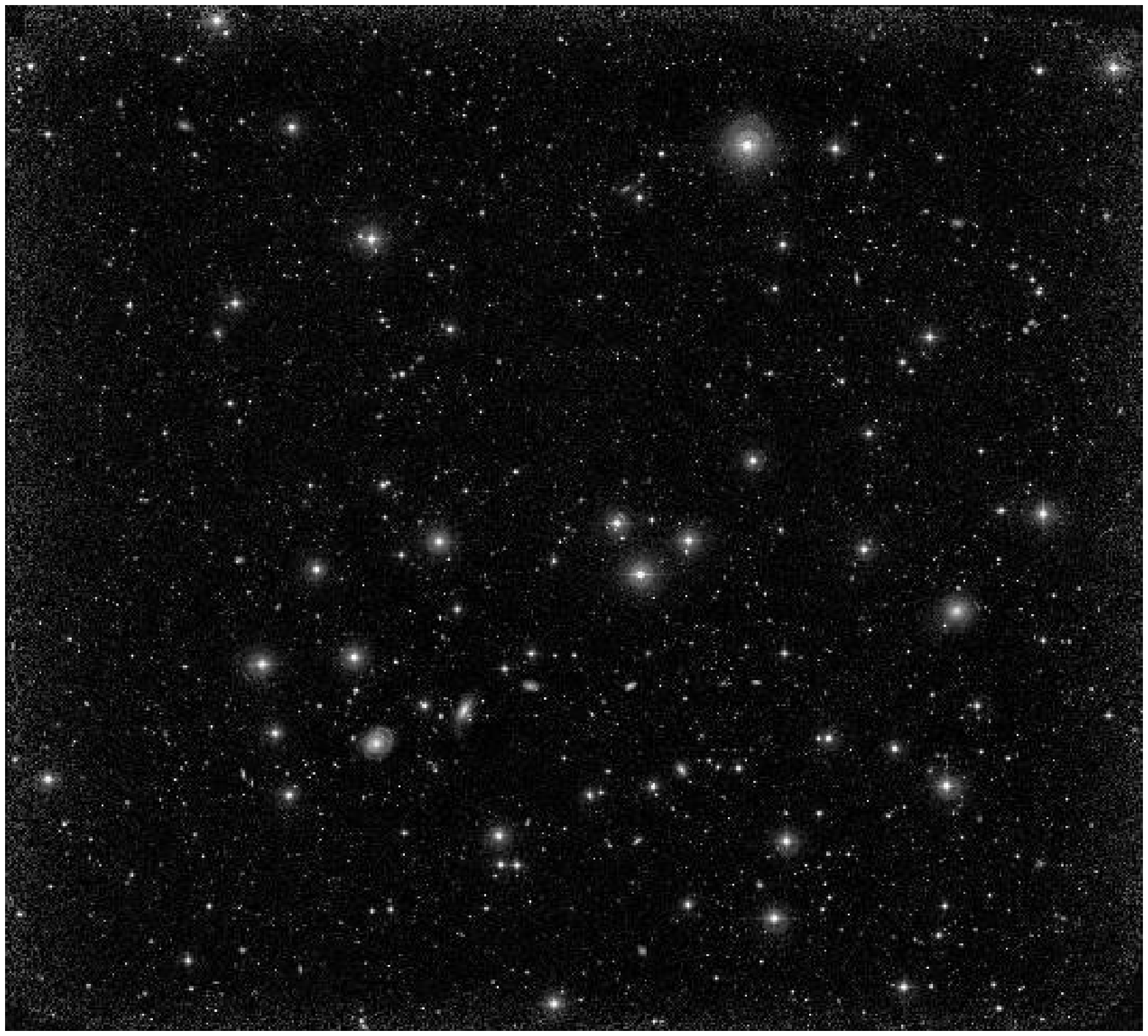}}
  \caption{Color composite $BVR$-image of the DEEP1b field with north
  up and east to the left. The image covers the entire WFI field of
  $34'\times33'$. It also demonstrates the accuracy of the
  astrometric calibration independently achieved in each passband.}
  \label{fig:dps-overview}
\end{figure*}

Final stacked images for each observed DPS field were produced from
nightly reduced images properly grouped in \emph{ Stacking Blocks}
(SBs) as explained in detail in Paper~I. The 304 reduced images with
grades better than D were converted into 42 stacked (co-added) images
out of which 40 were released and described in the present
paper. These stacks and their associated logs are publicly available
in the EIS web
pages.
Fig.~\ref{fig:dps-overview} shows a color composite image of the DEEP1b field typical for
the fields observed in this survey. These high-galactic latitude
fields contrast with those of Paper~I, which also included a number of
low-galactic latitude fields.

As before all images were visually inspected and graded. Out of the 42
stacked images covering the DPS fields, 30 were graded A, 6 B, 4 C,
and 2 D. When appropriate a comment is associated to the image.
Table~\ref{tab:stack-comments}
illustrates that by listing the 21 images for which comments were
made, regardless of their grade. 

\begin{table*}[htbp]
\center
\caption{Grades and comments for stacked images with an associated comment. 
The table, ordered by increasing
grade, lists: in Col.~1 the field
name; in Col.~2 the passband; in Col.~3 the grade; and in Col.4 the
associated comment.}
\label{tab:stack-comments}
\begin{tabular}{lccl}
\hline\hline
Field Name & Passband & Grade & Comment \\
\hline
DEEP3a&$I$\#845&D&fringing \\
DEEP3d&$B$\#842&D&single 300 sec image \\
DEEP1a&$I$\#845&C&fringing \\
DEEP2b&$I$\#845&C&fringing \\
DEEP2c&$I$\#845&C&fringing \\
DEEP3c&$I$\#845&C&fringing - Poor background subtraction near bright stars \\
DEEP1a&$U$\#841&B&visible electronic noise \\
DEEP1b&$U$\#877&B&visible electronic noise \\
DEEP2c&$U$\#841&B&visible electronic noise \\
DEEP3a&$U$\#841&B&visible electronic noise \\
DEEP3a&$U$\#877&B&visible electronic noise \\
DEEP2d&$R$\#844&B&Reflections, few images stacked \\
DEEP3b&$I$\#845&B&fringing near bright star in lower left (SE) \\
DEEP1b&$I$\#845&A&low level fringing \\
DEEP2a&$R$\#844&A&poor background subtraction around the bright star on the lower left corner \\
DEEP2b&$V$\#843&A&residual traces of the inter-chip gaps visible \\
DEEP3a&$R$\#844&A&stray light reflections in the upper and lower left corners \\
DEEP3a&$V$\#843&A&stray light reflections in the upper and lower left corners \\
DEEP3a&$B$\#842&A&stray light reflections in the upper and lower left corners \\
DEEP3c&$R$\#844&A&poor background subtraction near bright stars  \\
DEEP3d&$I$\#879&A&low level fringing - stray light reflection on the lower right corner \\
DEEP3d&$V$\#843&A&stray light reflection on the lower right corner  \\
\hline
\end{tabular}
\end{table*}

Of the two images graded ``D'', one, the $B$-band image of DEEP3d, was discarded because it was a
single image of 5 minutes exposure time, and the other, an $I$-band image of DEEP3a produced
from a single RB comprising 6 exposures, because it had a very strong
fringing pattern.

Table~\ref{tab:gradepass} shows for each filter the grade
breakdown for the stacked images. Not surprisingly, comparison with
Table~\ref{tab:grade-distribution} shows that, in general, the grades
of stacks are considerably better than those of the reduced images. In
particular, in contrast to the reduced images, the $U$-bands are
mostly graded A.  The notable exception are the $I$-band (\#845)
images, of which a significant fraction are graded C, in
fact all images graded C are $I$-band, for which the procedure of
de-fringing was less than ideal.  This is in marked contrast with the
results reported in Paper I
for the XMM survey,
indicating the importance of the observing strategy for properly
correcting for fringing. The results also show that fringing is the
single most important problem faceed by WFI multi-wavelength surveys.

\begin{table}
\center
\caption{Grades assigned to the DPS stacked images as a function of
the passband. The table gives: in Col.~1 the
passband; in Col.~2 the ESO filter number; in Col.~3
the total number of stacked images for each passband; in Cols.~4--7
the ratio of images with a given grade to the total number of images
taken with that filter.}
\label{tab:gradepass}
\begin{tabular}{ccccccc}
\hline\hline
Passband & Filter & \#Stacks & A & B & C & D \\
\hline
$U$&\#877&7&0.71&0.29&0.0&0.0\\
$U$&\#841&3&0.33&0.67&0.0&0.0\\
$B$&\#842&7&0.86&0.0&0.0&0.14\\
$V$&\#843&8&1.0&0.0&0.0&0.0\\
$R$&\#844&9&0.88&0.12&0.0&0.0\\
$I$&\#879&1&1.0&0.0&0.0&0.0\\
$I$&\#845&7&0.14&0.14&0.57&0.14\\
\hline
\end{tabular}
\end{table}

The accuracy of the final photometric calibration depends
on the accuracy of the photometric calibration of the reduced images
which are used to produce the final co-added stacks and the number of
independent photometric nights in which these were observed.           
The former depends not only on the quality of
the night but also on the adopted calibration plan. To preview the
quality of the photometric calibration, Table~\ref{tab:images}
provides information on the number of reduced images and number of
independent nights for each passband and filter.    Examining this table together with
Table~\ref{tab:field-calib} provides some insight into the quality of
the photometric calibration of each final stack.

\begin{table*}
\center
\caption{Number of reduced images and number of independent nights for
each field and passband. The table gives for
each field: in Col.~1 the region;
in Col.~2 the field name; in Cols.~3--9 the number of reduced images,
and in parenthesis, the number of independent nights for each
passband.}
\label{tab:images}
\begin{tabular}{llccccccc}
\hline\hline
Region & Field Name & $U$\#877&$U$\#841&$B$\#842&$V$\#843&$R$\#844&$I$\#879&$I$\#845 \\
\hline
DEEP1&DEEP1a & 20 (12)  & 5 (2)  & 4 (1)  & 0 (0)  & 4 (2)  & 0 (0)  & 10 (4)  \\
DEEP1&DEEP1b & 11 (9)  & 0 (0)  & 6 (4)  & 5 (5)  & 11 (4)  & 0 (0)  & 11 (6)  \\
DEEP1&DEEP1c & 0 (0)  & 0 (0)  & 0 (0)  & 3 (3)  & 6 (4)  & 0 (0)  & 0 (0)  \\
DEEP2&DEEP2a & 0 (0)  & 0 (0)  & 0 (0)  & 0 (0)  & 6 (3)  & 0 (0)  & 0 (0)  \\
DEEP2&DEEP2b & 17 (5)  & 0 (0)  & 7 (2)  & 5 (2)  & 7 (3)  & 0 (0)  & 8 (5)  \\
DEEP2&DEEP2c & 12 (5)  & 9 (3)  & 0 (0)  & 7 (3)  & 0 (0)  & 0 (0)  & 12 (3)  \\
DEEP2&DEEP2d & 0 (0)  & 0 (0)  & 0 (0)  & 0 (0)  & 2 (2)  & 0 (0)  & 0 (0)  \\
DEEP3&DEEP3a & 10 (4)  & 7 (4)  & 9 (6)  & 5 (5)  & 6 (5)  & 0 (0)  & 1 (1)  \\
DEEP3&DEEP3b & 9 (3)  & 0 (0)  & 6 (2)  & 4 (2)  & 2 (1)  & 0 (0)  & 7 (5)  \\
DEEP3&DEEP3c & 18 (7)  & 0 (0)  & 5 (2)  & 3 (2)  & 4 (3)  & 0 (0)  & 8 (6)  \\
DEEP3&DEEP3d & 0 (0)  & 0 (0)  & 1 (1)  & 3 (2)  & 0 (0)  & 8 (5)  & 0 (0)  \\
\hline
\end{tabular}
\end{table*}

The main attributes of the stacks produced for each field/filter
combination are given in Table~\ref{tab:attributes}. The table can also be used to assess
the completeness of the survey as a whole.

To provide a backbone infrastructure to allow the user to have an
overview of the quality of the released data, an integral part of the
data release infrastructure provided by the EIS Data Reduction System
is to produce figures illustrating the distribution of attributes of
the released products. These distributions for the data considered are
made available in the EIS web
pages.
as part of the data release.  The figures include: the distribution of
grades, limiting magnitude (Vega, 5 sigma, 2 \arcsec aperture), seeing
and an estimate of the rms of the PSF distortions; and the relation of
limiting magnitude (Vega, 5 sigma, 2 \arcsec aperture) versus seeing
and integration time.  For the sake of space these plots are not
reproduced here, even though the present release may differ slightly
from that shown in the web.

\begin{table*}[ht]
  \centering
  \caption{Overview of the attributes of the produced image
    stacks and survey completeness. The table gives:
in Col.~1 the field name; in Col.~2 the passband; in Col.~3 the filter
identification; in Col.~4 the total integration time $T_{int}$ in
seconds, of the final stack; in Col.~5 the number of contributing
reduced images or RBs; in Col.~6 the total number of science frames
contributing to the final stack; in Cols.~7 and 8 the seeing in
arcseconds and the point-spread function (PSF) anisotropy measured in
the final stack; in Col.~9 the limiting magnitude, $m_{lim}$,
estimated for the final image stack for a 2\arcsec aperture, $5\sigma$
detection limit in the Vega system; in Col.~10 the grade assigned to
the final image during visual inspection (ranging from A to D); in
Col.~11 the fraction (in percentage) of observing time relative to
that originally planned.}
  \label{tab:attributes}
  \begin{tabular}{llrrrrrrrrr}
    \hline\hline
    Field & Passband & Filter & $T_{\rm int}$  & \#RBs & \#Exp. &
    Seeing & PSF rms & $m_\mathrm{lim}$  &Grade &
    Completeness\\ 
     &         &     & (sec)&        &  & (arcsec) &  & (Vega mag) &  & (\%) \\
\hline
DEEP1a& $U$&\#877&77393& 20  &86 &1.37 & 0.217 & 25.26 &A&179 \\
DEEP1a& $U$&\#841&18298& 5   &21 &1.28 & 0.086 & 24.42 &B& 30 \\
DEEP1a& $B$&\#842&11397& 4   &38 &1.37 & 0.139 & 25.85 &A& 90 \\
DEEP1a& $R$&\#844&9897&  4   &33 &0.87 & 0.169 & 25.74 &A&110 \\
DEEP1a& $I$&\#845&29693& 10  &83 &0.86 & 0.217 & 23.76 &C&110 \\
\hline                         
DEEP1b& $U$&\#877&27546& 11  &45 &1.17&0.235   & 24.62&B& 64 \\
DEEP1b& $B$&\#842&11397&  6  &38 &1.43&0.139   & 25.66&A& 90 \\
DEEP1b& $V$&\#843&10497&  5  &35 &1.31&0.231   & 25.35&A&117 \\
DEEP1b& $R$&\#844&20694& 11  &69 &1.29&0.173   & 25.32&A&230 \\
DEEP1b& $I$&\#845&28492& 11  &95 &0.97&0.229   & 24.19&A&106 \\
\hline                         
DEEP1c& $V$&\#843&7498&   3   &25 &1.19&0.147   & 25.03&A& 83 \\
DEEP1c& $R$&\#844&11997& 6  &40 &0.98&0.147   & 25.43&A&133 \\
\hline                         
DEEP2a& $R$&\#844&5998&   6  &20 &0.84&0.105   & 24.51&A& 67 \\
\hline                         
DEEP2b& $U$&\#877&55795& 15  &62 &1.2&0.251   & 23.87&A&129 \\
DEEP2b& $B$&\#842&11997&  7  &40 &0.98&0.212   & 26.48&A& 95 \\
DEEP2b& $V$&\#843&9297 &  5  &31 &0.89 &0.189   & 25.19&A&103 \\
DEEP2b& $R$&\#844&11997&  7  &40 &1.36&0.127   & 25.06&A&133 \\
DEEP2b& $I$&\#845&17697&  8   &42 &0.81&0.146   & 23.77 &C& 66 \\
\hline                         
DEEP2c& $U$&\#877&45896& 12  &51& 2.26 &0.265   & 25.05&A  &106 \\
DEEP2c& $U$&\#841&32397&  9  &36& 0.94&0.332   & 24.56&A  & 53 \\
DEEP2c& $V$&\#843&16597&  7  &39& 1.07&0.218   & 24.94&A  &184 \\
DEEP2c& $I$&\#845&29695&  9  &50& 0.91&0.155   & 23.94&C &110 \\
\hline                         
DEEP2d& $R$&\#844&600&   1   &2& 1.00&0.096   & 23.74&B& 7 \\
\hline                          
DEEP3a& $U$&\#877&35997& 10 &40& 1.12&0.223   & 24.29&B&  83 \\
DEEP3a& $U$&\#841&27897&  7  &31& 1.28&0.159   & 24.11&B& 46 \\
DEEP3a& $B$&\#842&11997&  9  &40& 0.99&0.178   & 27,73&A& 95 \\
DEEP3a& $V$&\#843&8997 &  5  &30& 1.03&0.171   & 25.1 &A&100 \\
DEEP3a& $R$&\#844&9597 &  6  &32& 0.82&0.158   & 24.99&A&107 \\
DEEP3a& $I$&\#845&3599 &  1  &6 & 1.50&0.047  & 22.3 &D& 13 \\
\hline                         
DEEP3b& $U$&\#877&35997&  9  &40& 0.96&0.224   & 24.39&A& 83 \\
DEEP3b& $B$&\#842&11697&  6  &39& 0.96&0,211   & 27.2 &A& 93 \\
DEEP3b& $V$&\#843&8997 &  4  &30& 0.89&0.157   & 26.68&A&100 \\
DEEP3b& $R$&\#844&5998 &  2  &20& 0.78&0.096   & 24.73&A& 67 \\
DEEP3b& $I$&\#845&20995&  7  &65& 0.84&0.155   & 23.35&B& 78 \\
\hline                         
DEEP3c& $U$&\#877&48844&  18  &64& 0.99&0.266   & 23.89&A&113 \\
DEEP3c& $B$&\#842&13496&  5   &45& 0.93&0.182   & 25.83&A&107 \\
DEEP3c& $V$&\#843&5998 &  3   &20& 0.79&0.121   & 25.07&A& 67 \\
DEEP3c& $R$&\#844&8997 &  4   &30& 0.81&0.154   & 25.43&A&100 \\
DEEP3c& $I$&\#845&23993&  8   &80& 0.98&0.182   & 23.85&C& 89 \\
\hline                          
DEEP3d& $B$&\#842&300  &  1 &1 & 1.58&0.072   & 25.32&D& 2  \\
DEEP3d& $V$&\#843&5998 &  3 &20& 0.95&0.107   & 24.75&A& 67 \\
DEEP3d& $I$&\#879&23193&  8 &79& 0.78&0.178   & 24.32&A& 86 \\
\hline
\end{tabular}
\end{table*}

\subsection {Source Catalogs}

Catalogs were extracted from the image stacks using the SExtractor
program (version 2.3.2, 
\citealt{Bertin1996}) and a common configuration file, with the
option of using the weight-map associated to each image. 
Details about the choice of input parameters and their fine-tuning can
be found in Paper~I.

Before being released the catalogs were superposed onto the
images, examined by eye and graded. The grade is meant to be a
subjective assessment of the overall appearance of the catalog in
terms of spurious objects, automatic masking of satellite tracks and
bright objects. To some extent the catalog grade reflects the quality
of the image. Information regarding the scientific quality of the
catalog can be found in the verification section of the associated
product log.  Out of the 40 catalogs released, 8 were graded A, 23 B,
7 C, and 2 D. In contrast to the images, the two catalogs graded D are
also being released in order to illustrate the impact of inadequate
de-fringing. It is important to note that the catalogs graded D were
extracted from grade C images.  Table~\ref{tab:grade-cat} presents all
34 cases where comments were made.

\begin{table*}
\center
\caption {Grades for source lists with associated comment. The table, ordered by increasing
grade, lists: in Col.~1 the EIS field name; in Col.~2 the passband; in
Col.~3 the grade; and in Col.~4 the associated comment.}
\label{tab:grade-cat}
\begin{tabular}{lccl}
\hline\hline
Field Name & Passband & Grade & Comment \\
\hline
DEEP1a&$I$\#845&D&Fringing leads to multiple spurious detections \\
DEEP2b&$I$\#845&D&Fringing leads to multiple spurious detections \\
DEEP2a&$R$\#844&C&Large number of spurious objects caused by very bright star at the lower left\\
& & & edge of the image. Missing masks around a few saturated stars \\
DEEP2c&$I$\#845&C&Reflections from bright stars cause spurious objects in various areas of the image.\\
& & &  Low level fringing adds also additional noise in the catalog \\
DEEP2d&$R$\#844&C&Numerous cosmic rays misidentified as real objects. Spurious objects caused by \\
& & & reflection ring of bright star in the lower left quadrant.  \\
DEEP3b&$I$\#845&C&Multiple spurious detections at the lower left quadrant due to fringing \\
DEEP3c&$R$\#844&C&Missing masks for a few saturated stars. Multiple spurious objects caused by \\
& &  & the reflection rings of the bright stars at the top and bottom right of the image \\
DEEP3c&$I$\#845&C&Spurious objects caused by fringing at the lower left quadrant of the image. \\
& & & Multiple spurious objects caused by the reflection rings of bright stars\\
& & &  at the top and bottom right of the image. \\
DEEP3d&$I$\#879&C&Spurious objects caused by multiples reflection rings/stray light across the image.  \\
DEEP1a&$U$\#877&B&Spurious objects around bright stars \\
DEEP1a&$U$\#841&B&Few spurious objects around bright stars \\
DEEP1a&$B$\#842&B&Mask missing for a few saturated stars \\
DEEP1b&$U$\#877&B&Masks missing around saturated stars. Spurious objects around bright stars. \\
DEEP1b&$I$\#845&B&Residual fringing increases the number of spurious objects\\
DEEP1c&$V$\#843&B&Masks missing around a few saturated stars \\
DEEP1c&$R$\#844&B&Masks missing around a few saturated stars \\
DEEP2b&$U$\#877&B&Masks missing around all saturated stars \\
DEEP2b&$B$\#842&B&Spurious objects caused by reflections of bright star at center top of the image. \\
& & & Missing masks for several saturated stars \\
DEEP2b&$V$\#843&B&Spurious objects caused by reflections of bright stars at the top center \\
& & &  and lower right corner of the image.  \\
DEEP2c&$U$\#877&B&Missing masks for a few saturated stars \\
DEEP2c&$U$\#841&B&Missing masks around a few saturated stars \\
DEEP3a&$U$\#877&B&Spurious objects around bright stars. Spurious objects caused by stray light \\
& & & at the top left corner. \\
DEEP3a&$U$\#841&B&Spurious objects around bright stars. Spurious objects caused by stray light\\
& & &  at the top left corner. \\
DEEP3a&$B$\#842&B&Masks missing for a few saturated stars. Spurious objects caused by stray light\\
& & &  at the top left corner. \\
DEEP3a&$V$\#843&B&Masks missing for a few saturated stars. Spurious objects caused by stray light \\
& & & at the top left corner \\
DEEP3a&$R$\#844&B&Masks missing for a few saturated stars. Spurious objects caused by stray light \\
& & & at the top and bottom left corners. \\
DEEP3b&$U$\#877&B&Spurious objects around bright stars \\
DEEP3b&$B$\#842&B&Missing masks around a few saturated stars \\
DEEP3b&$V$\#843&B&Masks missing around a few saturated stars \\
DEEP3b&$R$\#844&B&Spurious objects caused by left-over cosmic rays at the inter-chip gaps \\
DEEP3c&$U$\#877&B&Spurious objects around bright stars \\
DEEP3c&$V$\#843&B&Multiple spurious objects caused by the reflection ring of the bright stars \\
& & & at the top and bottom right of the image \\
DEEP1b&$R$\#844&A&Mask missing around 1 saturated star at the center-right of the image \\
DEEP2b&$R$\#844&A&Few spurious objects at the lower right corner caused by bright star reflection at the edge of the image. \\
\hline
\end{tabular}
\end{table*}

In general, there is a correlation between the grade of the catalogs
and the grade of the images from which they were extracted. 
However, the catalog grade can be lower than that of the associated
image due to effects such as the detection of large number of
spurious objects associated with the ghost images of bright stars,
stray light, and others. For instance, all the $U$-band catalogs were
graded B. As in the case of the corresponding images, the $I$-band
catalogs are typically C and D. The following two cases should be
noted:
   
\begin{enumerate}

\item DEEP1a (I) - the original image was graded C because of observed
  fringing in the final stack.  This has led to a catalog with no
  scientific value because of the high number of spurious objects
  detected along the fringing pattern. This catalog is being released
  {\it exclusively} as an illustration.

\item DEEP2b (I) - same as above

\end{enumerate}

The inspection of the projected distribution of objects,
strongly suggests that the automatic masking of satellite tracks has
worked remarkably well as no prominent linear features, a signature of
this type of problem, are seen on the inspected catalogs.
A description of the catalog format is found in the appendices of 
Paper~I.

\section {Discussion}
\label{sec:discussion}

\subsection{Comparison of counts and colors}

A key element in public surveys is to provide potential users with
information regarding the quality of the products released. To this
end a number of checks of the data are carried out and several
diagnostics plots summarizing the results are automatically produced
by the EIS Survey System. They are an integral part of the product
logs available from the survey release page. Due to the large number
of plots produced in the verification process these are not reproduced
here.  Instead, to illustrate the results only a small set of
these plots are shown.

A relatively simple statistics that can be used to check the catalogs
and the star/galaxy separation criteria is to compare the star and
galaxy number counts derived from the data to that of other authors
and/or to model predictions. For stars we use 
Vega magnitudes and for galaxies AB magnitydes, while for both we use the
MAG\_AUTO magnitudes from the catalog.
Figs.~\ref{fig:gal_deep1}--\ref{fig:gal_deep3} in
appendix~\ref{app:galcounts} show for each region a mosaic where the
rows represent the passbands from $U$(top) to $I$ (bottom) and the
columns the different fields available from a (left) to d
(right). Here objects with CLASS\_STAR$<$0.95 were considered to be
galaxies.  Note that the number counts shown in the figure take into
account the effective covered area, which is available in the {\tt
FIELDS} table (see appendix~D of Paper~I). Inspection of all figures
shows that the galaxy counts obtained from the final produced stacks
are, in general, in excellent agreement with those obtained by
previous authors, \citep{Arnouts2001,Metcalfe2001}, regardless of
the passband. The figures also highlight some of the
challenges in dealing with mosaics, as the completeness of the counts
can be seen to vary from field to field.

A complementary test is to compare the stellar counts to those
predicted by models such as the galactic model of \cite{Girardi2005}.
Figs.~\ref{fig:starcounts1}--\ref{fig:starcounts3} in
appendix~\ref{app:starcounts} show these comparisons, in the same
format as the galaxy counts, for each of the observed regions. The
good agreement between the data and the model predictions is indeed
remarkable, especially considering the uncertainties in the zero-point
and the model. These results give further support to the photometric
calibration and the scientific quality of the science-grade data
produced automatically by the EIS system.

While useful to detect gross errors, number counts are not
sufficiently sensitive to identify more subtle differences. The
comparison of expected colors of stars with theoretical models
provides a better test of the accuracy of the photometric calibration
in the different bands. Using the model of \cite{Girardi2005},
the predicted colors of stars can be obtained and compared to the data
to evaluate at least the rough consistency of the photometric
calibration in different passbands.  Such comparisons were made for
each of the four fields covered in all five passbands and are
shown in Figs.~\ref{fig:starcol1}--\ref{fig:starcol4} presented in
appendix~\ref{app:colorcolor}. From these figures one finds that at
least for the fields with complete color information there is a fairly
good agreement between the locus of stars and those predicted by the
model in four different color-color diagrams, indicating a good
($\lesssim 0.1~mag$) consistency among the photometric calibration
obtained in different passbands.

\void{
Overall, the offsets from the
predicted colors are of the order that one can expect for automatic
calibrations as those done here and using the standard calibration
plan of ESO.\\
The $U$- and $B$-bands are the ones for which the largest offsets, up
to 0.2~mag in the worst case (DEEP2b), are observed. The redder bands,
in particular $R$ and $I$, have much smaller offsets which in all
cases are less than 0.1~mag and in many cases negligible. These
results are similar to those obtained for the XMM fields
\citep{Dietrich2005} and seem to indicate that the automatic
calibration and the standard calibration plan are better adapted to
the red than the blue bands.
The previous comparison of the expected and measured colors of the
stars, points out that in the case of long-term projects like those
carried out by EIS, it is important to have a careful calibration plan
if a high-precision absolute calibration of the data is required. The
most efficient way would probably be to assure, for each
field and passband combination, calibration data taken in a perfectly
photometric night. Such observations had to be accompanied by a
thorough set of standard star observations with a good coverage in
both airmass and color in order to obtain precise photometric
calibrations.
}

\void{ In the current project such data were not obtained, as
reflected by the small zero-point offsets.}

\subsection{Comparison with other reductions}

The data presented here have been reduced using the same method and
parameter settings as those discussed in Paper~I. In that paper the
results of the reduction through the EIS data reduction system and
that using the GaBoDS pipeline \citep{Schirmer2003,Erben2005}
 developed in Bonn were compared. Extensive and repeated
comparisons were made between SExtractor-produced catalogs from the
images generated by these two independent systems.  Initial
discrepancies, resulting from the different techniques used (e.g. 
cosmic ray removal, gain-harmonization) were resolved, leading at the
end to results in excellent agreement. 

A similar comparison has been partially carried out for the
present dataset and it has shown an important short-coming of the
current implementation of the EIS/MVM library. Even though
the chip-to-chip gain correction is not mandatory, the log file
created by the library provides no information of whether or not it
has been applied.  In fact, comparison of a subset of frames showed
that in the original reductions made public late 2004 the images for
DEEP2 were not homogenized. Since in the final, sky-subtracted image,
residual differences are not seen this was only detected from the
comparison with the independent reduction conducted in
Bonn (see also \citealt{Hildebrandt}) .  It is important to emphasize that the images for DEEP2 have
been re-reduced and have replaced those released in October 2004. The
same is true for the associated science-grade catalogs.

\void{
 One of the important implications of this is that the log
file produced by the EIS/MVM programs needs to be improved. For
instance, from the comparison it became evident that in one of the
regions (DEEP2) the correction for the chip-to-chip variations had not
been applied. However, the log file does not report whether or not
this correction is applied, and without proper reporting it becomes
difficult after sky subtraction to devise a procedure to check for
this in a systematic way.}

\subsection {Radio/Optical correlation}
\label{sec:atesp}

\begin{figure}[t]
  \resizebox{\hsize}{!}
  {\includegraphics{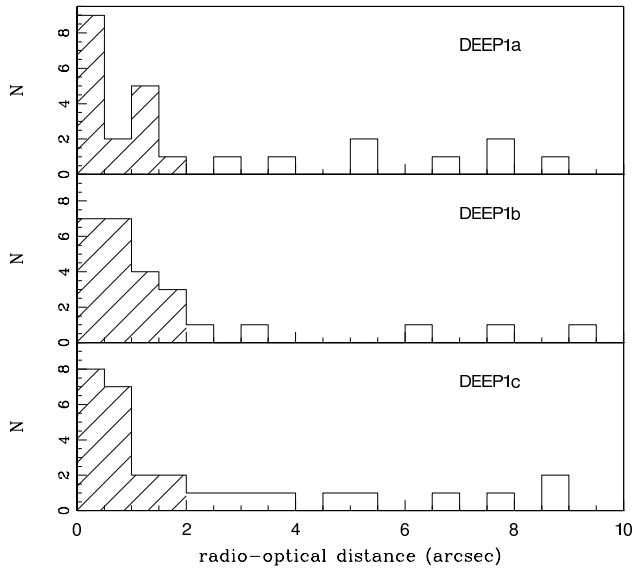}}
  \caption{ Distribution of separations bewtween radio sources and
the nearest $R$-detected sources for DEEP1a (upper panel), DEEP1b
(middle panel) and DEEP1c (lower panel).  The dashed histograms show
those considered as actual counterparts.}
\label{fig:atid}
\end{figure}


\begin{figure}[t]
  \resizebox{\hsize}{!}{\includegraphics{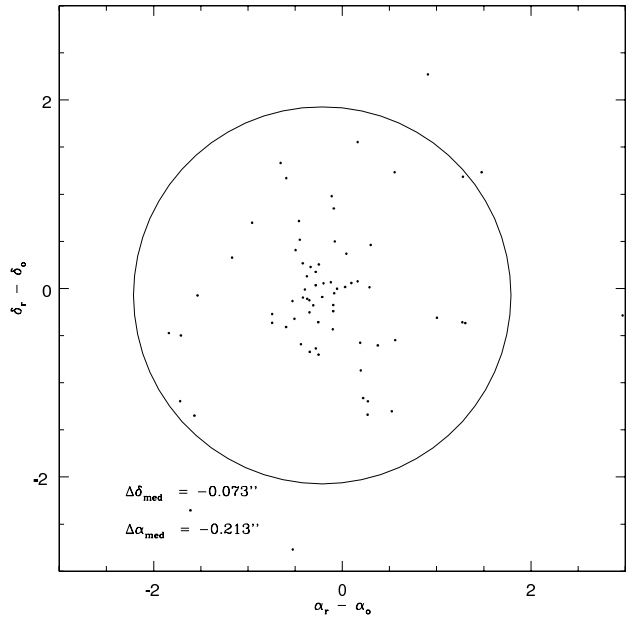}}
  \caption{Radio-optical offset computed for all the optical
 counterparts in the fields  DEEP1 a,b,c.  The 2 arcsec
 circle defines the boundaries of those considered to be ``good''
 identifications. Also reported are the values of the median of the
 offset distribution in right ascension and declination.}
\label{fig:offset}
\end{figure}

As mentioned earlier, the DEEP1 region was chosen to complement
the sub-mJy ATESP 1.4 GHz radio survey carried out with the Australia
Telescope Compact Array (ATCA). The ATESP survey covers a strip of
$26\times 1$ square~degrees, at declination $\delta\sim-40^\circ$, and
has provided a catalogue of $\sim3000$ radio sources to a flux limit
of $S_{1.4 GHz}\sim0.4$ mJy \citep{Prandoni2000a,Prandoni2000b}.  The
aim of the ATESP radio survey is to study the evolutionary properties
of the objects belonging to the mJy and sub-mJy populations, by
directly measuring the evolution of star-forming radio sources, in the
redshift range $0<z<1$, and of low-luminosity AGNs up to higher
redshifts.

In the $4\times0.25$~square degrees of the DEEP1 region covered by DPS
there are 109 ATESP 1.4 GHz radio sources - 25 in DEEP1a, 26 in
DEEP1b, 29 in DEEP1c and 29 in DEEP1d. Since the latter field was
not covered by the present survey, the analysis below refers to only
80 radio sources. To identify optical counterparts of these radio
sources the nearest optical object was searched for. All the available
single passband catalogs for the fields DEEP1a, b and c were treated
separately.  Fig.~\ref{fig:atid} shows the radio-optical distance
distribution of the $R$-selected optical counterparts of the
ATESP radio sources for DEEP1a, b and c, respectively. As can be
seen most optical counterparts are found within 2~arcsec. Beyond this
distance, the distribution is flat, as expected for spurious
identifications. Therefore, in this preliminary analysis this value is
taken to be the maximum distance from a radio source for an R-band
object to be considered its optical counterpart. The distribution of
offsets for all the identified radio source-optical pairs is shown in
Fig.~\ref{fig:offset}. It is worth noticing that systematic offsets in
the relative position between the radio and optical sources are only
$-0\farcs073$ in declination and $-0\farcs213$ in right ascension (see
median values in Fig.~\ref{fig:offset}), and have therefore been
neglected. Also note that matches obtained in the $R$-band are the
most complete, since all the radio sources identified in any of the
$U$-, $B$-, $V$- or $I$-bands are also identified in $R$-band.


The ATESP source identification rate varies from 66\% to 81\%,
depending on the field, while the estimated contamination  is 10-14\%. 
These identification rates
are consistent with the value of 74\% reported for the VVDS-VLA sample
\citep{Ciliegi2005}, where a radio/optical analysis of a sample of
$\mu$Jy sources was performed down to a similar optical depth.
It is also interesting to compare these results to those obtained
using a shallower survey, such as the $I-$band EIS-WIDE
survey \citep[Patch A, ][]{Nonino1999}. In the
3~square~degrees region that this patch covers, a radio/optical
analysis was carried out, yielding the identification of $\sim57\%$
ATESP radio sources down to $I=22.5$, the limiting magnitude of the optical
imaging \citep{Prandoni2001}. By comparing this identification rate
to the one obtained with the DPS data, one finds that there is a
significant increase, demonstrating the need for deep follow-up
surveys for properly identifying the mJy/sub-mJy population.

\begin{figure*}[t]
\resizebox{12cm}{!}{\includegraphics{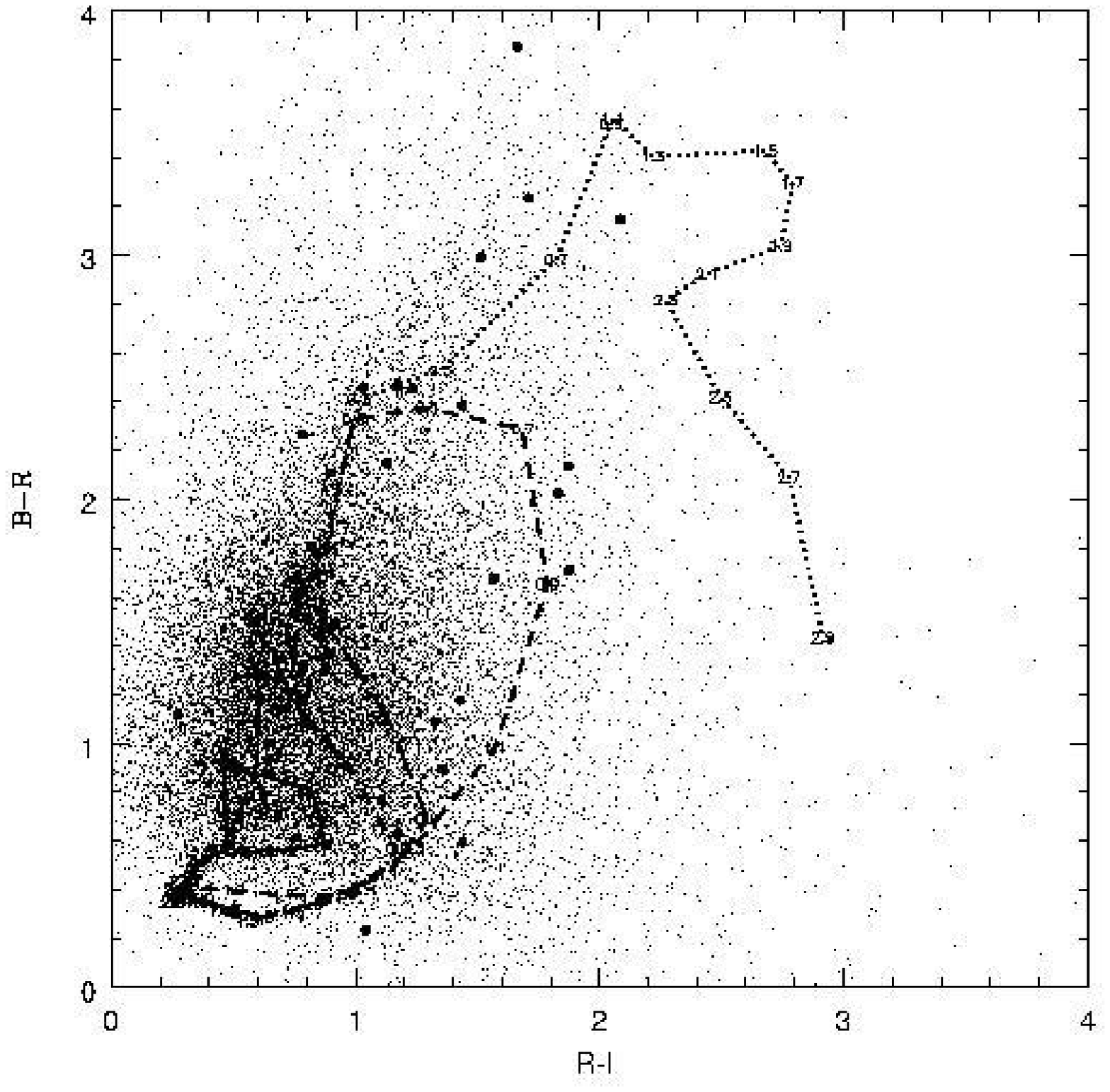}}
\hfill
\parbox[b]{55mm}{
\caption{Color-Color Diagram. In the figure is plotted the $(R-I)
\times (B-R)$ colors for the whole optical population (dots) and for
the optical counterparts of ATESP radio sources (filled circles). Dot,
dashed, long-dashed, dot-dashed and solid lines represent the
evolutionary tracks of bursts (single burst of star formation followed
by passive evolution), ellipticals, Sb and Sc spirals, and starburst
galaxies, respectively.  The evolutionary tracks were computed at
intervals of 0.2 in redshift using the GISSEL libraries
\citep{BruzualCharlot93}. The model tracks span the redshift range
$0.0<z<3.0$.}
\label{fig:atcol}}
\end{figure*}

The multi-band DPS data
was used to analyze the color properties of the ATESP radio sources.
Fig.~\ref{fig:atcol} shows in the ($R-I$)$\times$ ($B-R$) color diagram
the location of the optical objects associated to the radio sources
(filled circles), compared to the colors of field galaxies (dots).
Also plotted are the evolutionary tracks of various optical galaxy
populations (from early-type to spiral to starburst galaxies) in the
redshift range $0<z<3$.  From this plot it is clear that while the
majority of the ATESP sources follows the early-type galaxy tracks
(dotted and dashed lines) up to redshift $\sim2$,  some ATESP sources
have colors consistent with being spiral or starburst galaxies
(long-dashed, dot-dashed and solid lines) up to z$\sim0.5-1$.
These results are consistent with those of previous work
\citep[e.g. ][]{Prandoni2001} and the most likely interpretation is that
classical radio galaxies, usually associated to early-type objects
(ellipticals and/or S0) are, in general, more powerful and can be seen
to larger redshifts than radio sources associated to star-forming
galaxies.

To dramatically increase the identification rate of the ATESP sources,
one requires deep near-infrared ($J$, $H$, $K$) imaging. In this
respect, the infrared part of DPS \citep[see ][]{Olsen2006} is very
useful, since it provides $J$ and $K$ images down to $K_{AB}=21.3$
for the DEEP1a and b fields. Such data should allow the identification
of the most distant radio galaxies (reddened by redshift effects) and
possible ``rare'' intrinsically red objects (like EROs, ULIRGs, etc),
probably reddened by dust.

The preliminary radio/optical analysis described above will be further
developed in a future paper (Mignano {\it et~al.} in prep.), where the
multi-color information will be used to determine spectral types and
(photometric) redshifts for the identified radio sources.
We notice that, to fully exploit the capabilities of the optical data
described above, the entire DEEP1 $2^{\circ}\times 0.5^{\circ}$ region
has been imaged also at 5~GHz, again using the ATCA. The 5~GHz data
\citep{Prandoni2005} give the possibility of measuring the radio
spectral index (between 1.4 GHz and 5 GHz) of the radio sources. This
additional information may help constraining the origin
(star formation or nuclear activity) of the radio emission in mJy and
sub-mJy radio sources and in understanding whether it is linked to the
optical light.

\section {Summary}
\label{sec:summary}

This paper presents the survey products prepared by the EIS team
for the optical part of the Deep Public Survey (DPS) conducted over 4
years by the team.  The survey was designed to cover 3 independent
strips of the sky with an area of 1~square degree each in five
passbands. The paper describes the observations and presents a
comprehensive discussion of the set of products prepared and validated
by the EIS Data Reduction System.  These products are publicly
available including fully calibrated reduced images taken on a nightly
basis, final stacked images and extracted source catalogs. It should
be noted that the present optical data are complemented by
observations in infrared reported  by \cite{Olsen2006}.

\acknowledgements

We would like to express our thanks to several other EIS visitors who
have contributed directly or indirectly to the results presented in
this paper, too numerous to list individually.  Special thanks to
L. Girardi for providing us with the means to derive the color-color
tracks used in this paper, which were created for the filters used by
EIS. LFO acknowledges financial support from the Carlsberg Foundation,
the Danish Natural Science Research Council and the Poincar\'e
fellowship program at the Observatoire de la C\^ote d'Azur. The Dark
Cosmology Centre is funded by the Danish National Research Foundation.

\appendix

\section{Galaxy Counts}
\label{app:galcounts}

\begin{figure*}
\center
\resizebox{0.75\textwidth}{!}{\includegraphics{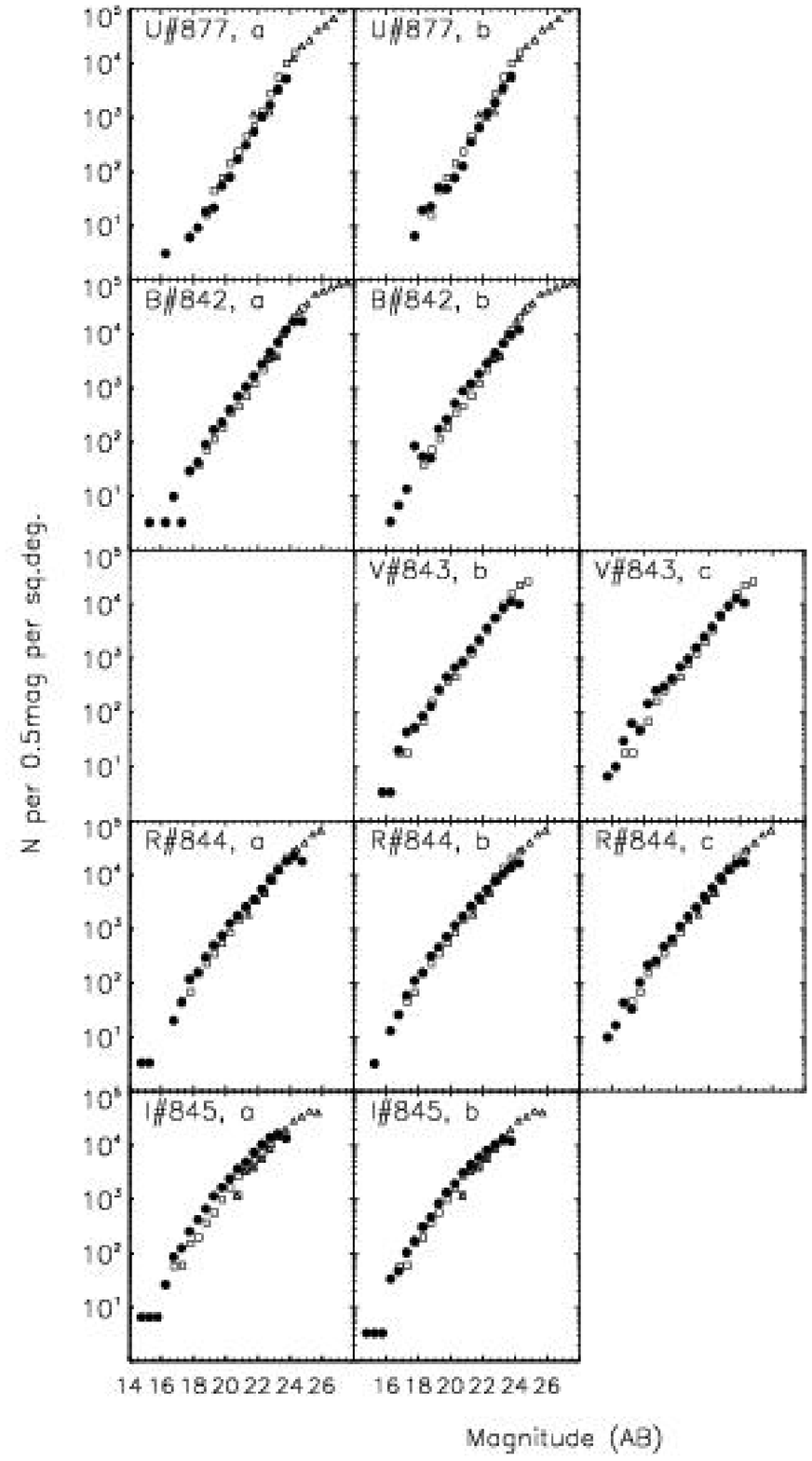}}
\caption{Galaxy number counts for the fields in the DEEP1 region
(solid points) compared to those of \cite{Arnouts2001} (open
squares) and  \cite{Metcalfe2001} (open triangles). The
passband, filter identification number and field are indicated in each
panel.}
\label{fig:gal_deep1}
\end{figure*}

\begin{figure*}
\center
\resizebox{0.75\textwidth}{!}{\includegraphics{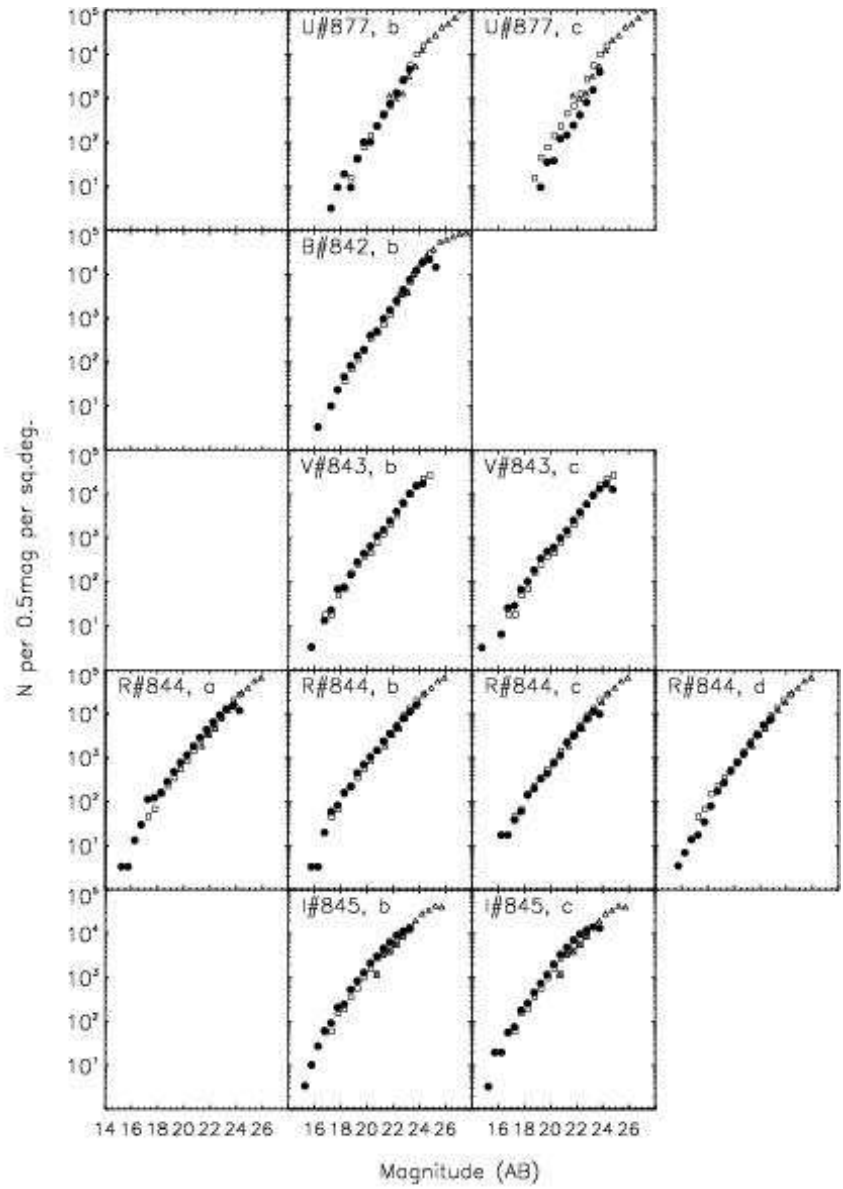}}
\caption{Same as in Fig.~\ref{fig:gal_deep1} for region Deep2. }
\label{fig:gal_deep2}
\end{figure*}

\begin{figure*}
\center
\resizebox{0.75\textwidth}{!}{\includegraphics{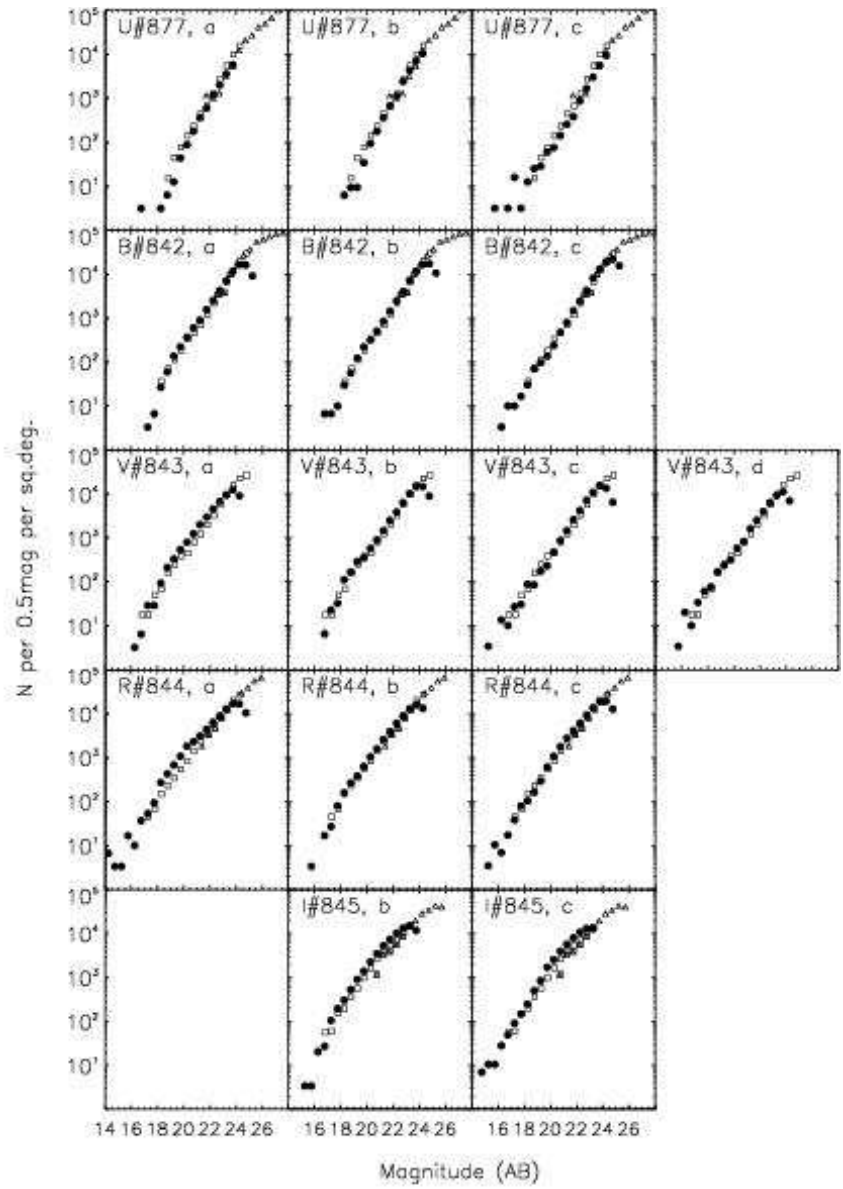}}
\caption{Same as in Fig.~\ref{fig:gal_deep1} for region Deep3. }
\label{fig:gal_deep3}
\end{figure*}

\section{Stellar Counts}
\label{app:starcounts}

\begin{figure*}
\center
\resizebox{0.75\textwidth}{!}{\includegraphics{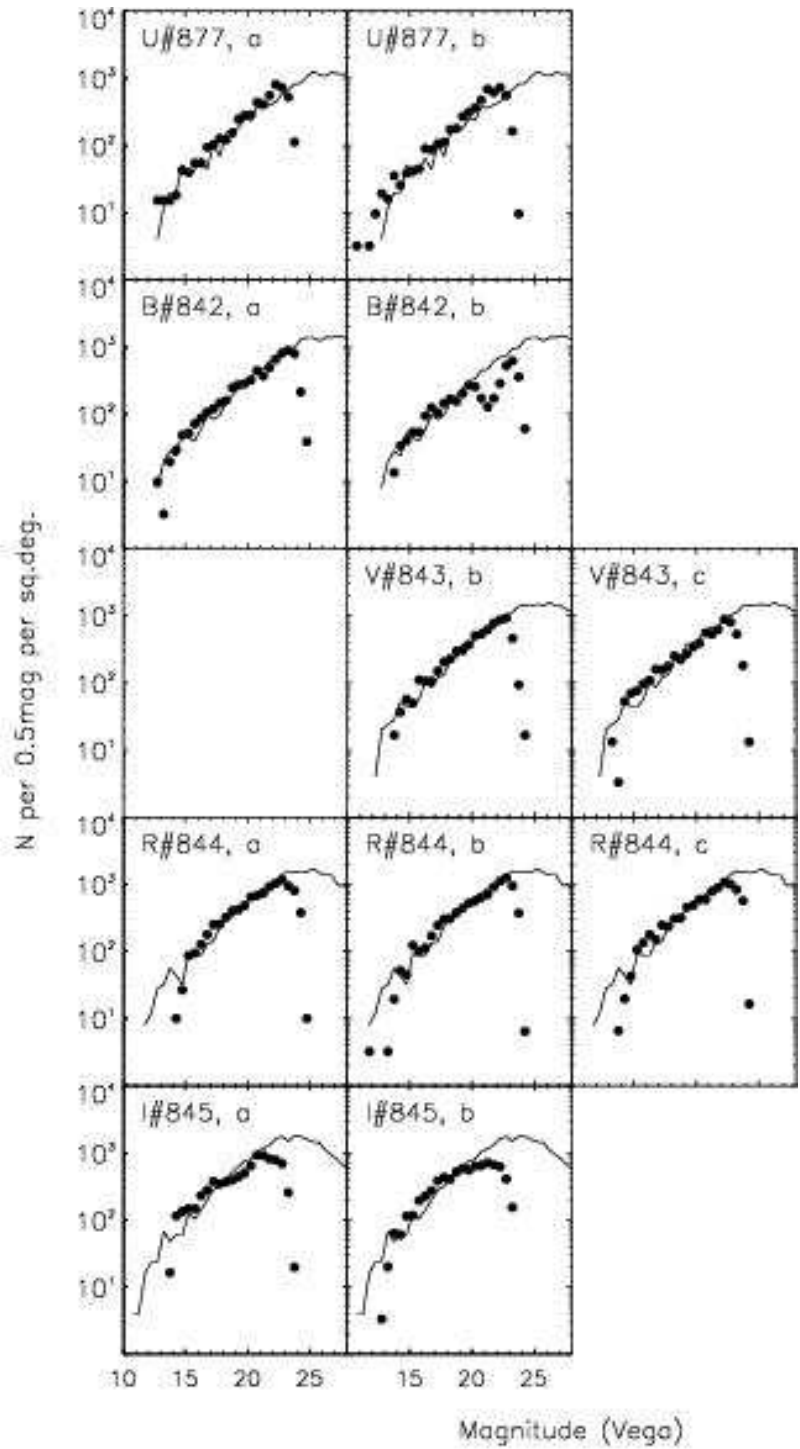}}
\caption{Stellar number counts (black points) compared with model
predictions (solid line) from \cite{Girardi2005}. The
passband, filter identification number and field are indicated in each
panel.}
\label{fig:starcounts1}
\end{figure*}

\begin{figure*}
\center
\resizebox{0.75\textwidth}{!}{\includegraphics{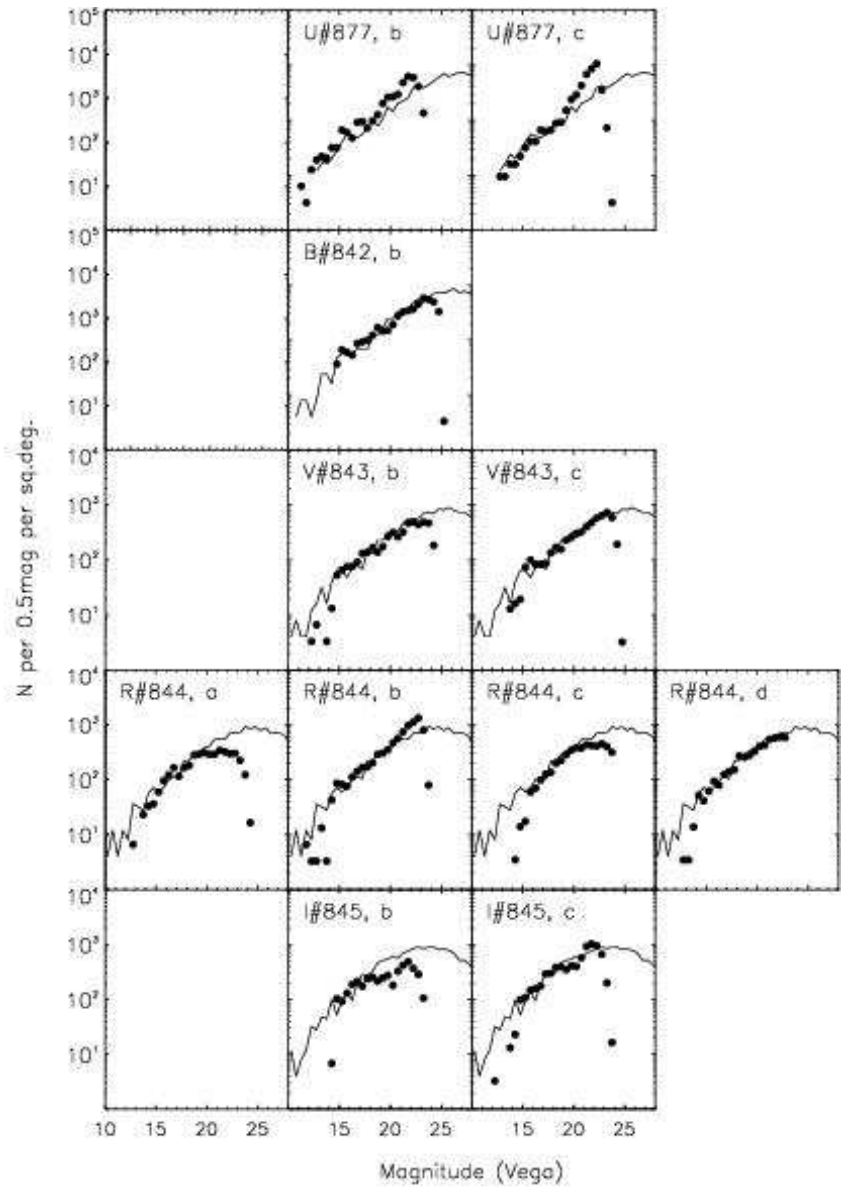}}
\caption{Same as Fig.~\ref{fig:starcounts1} but for the DEEP2 region.}
\label{fig:starcounts2}
\end{figure*}

\begin{figure*}
\center
\resizebox{0.75\textwidth}{!}{\includegraphics{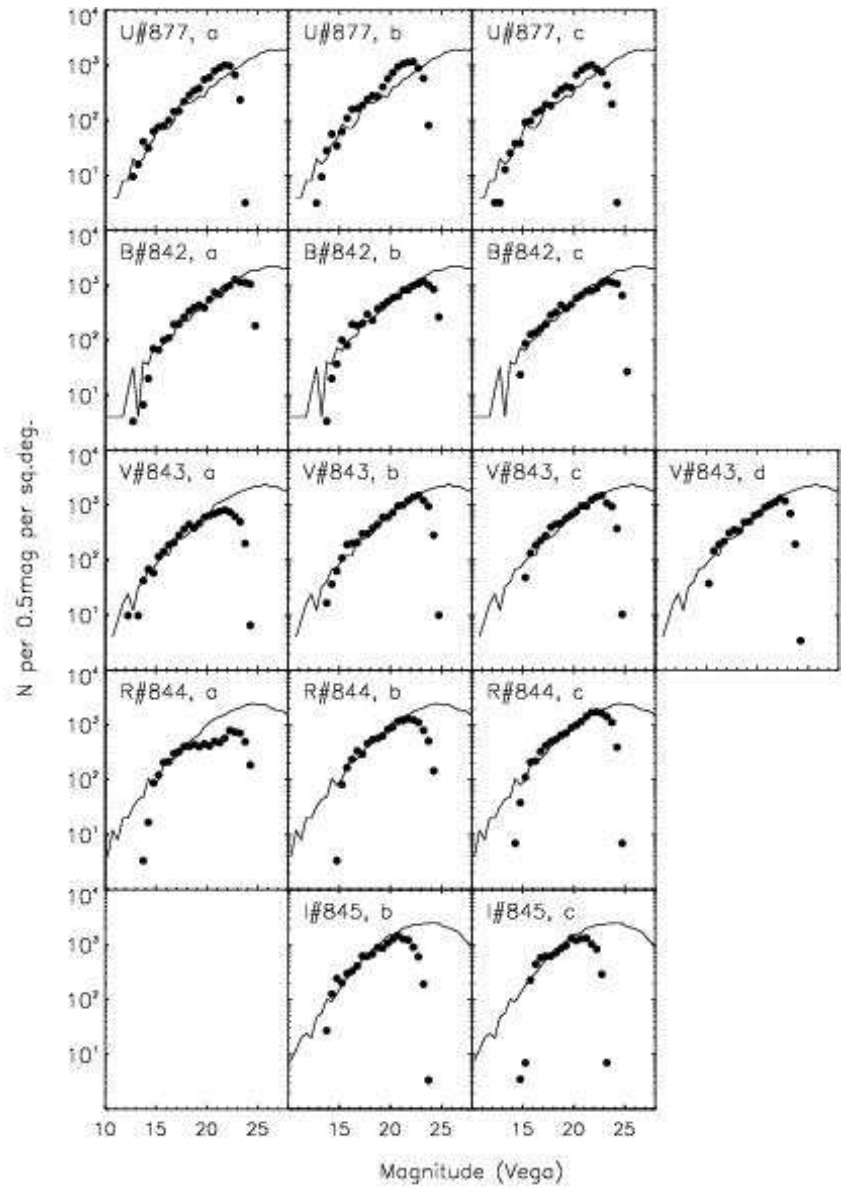}}
\caption{Same as Fig.~\ref{fig:starcounts1} but for the DEEP3 region.}
\label{fig:starcounts3}
\end{figure*}

\section{Color-color Diagrams}
\label{app:colorcolor}

\begin{figure*}
\center
\resizebox{0.4\textwidth}{!}{\includegraphics{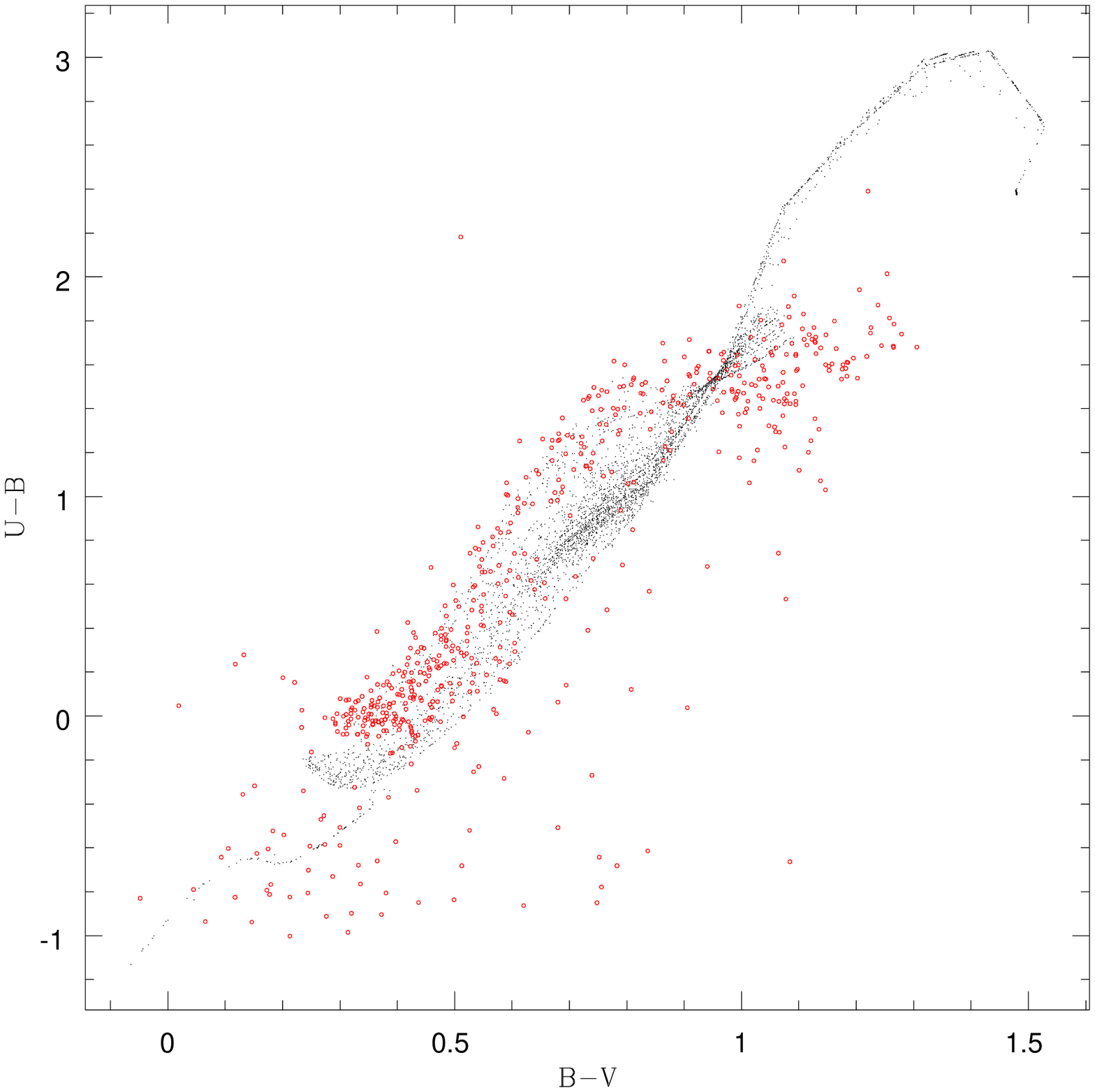}}
\resizebox{0.4\textwidth}{!}{\includegraphics{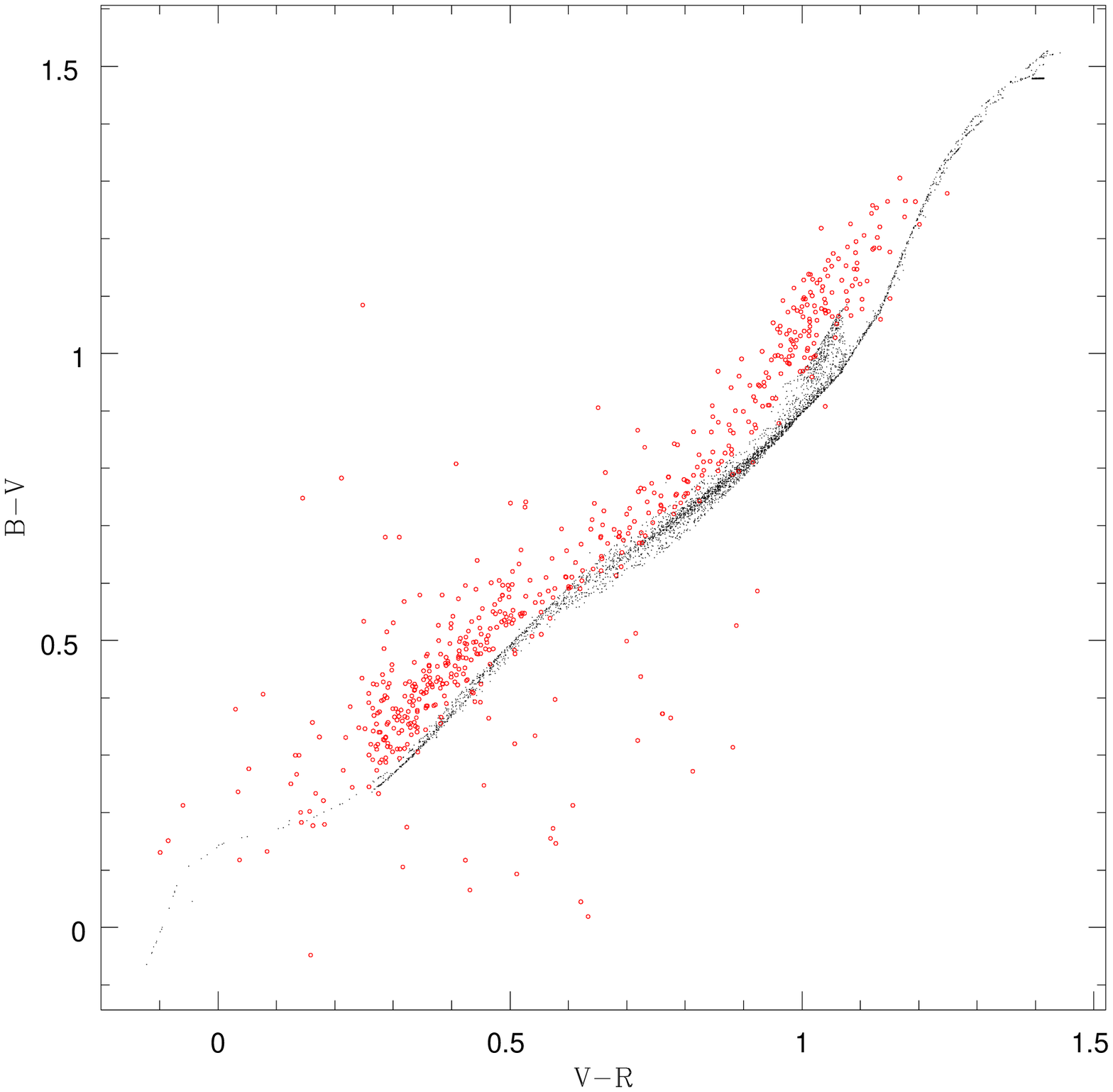}}

\resizebox{0.4\textwidth}{!}{\includegraphics{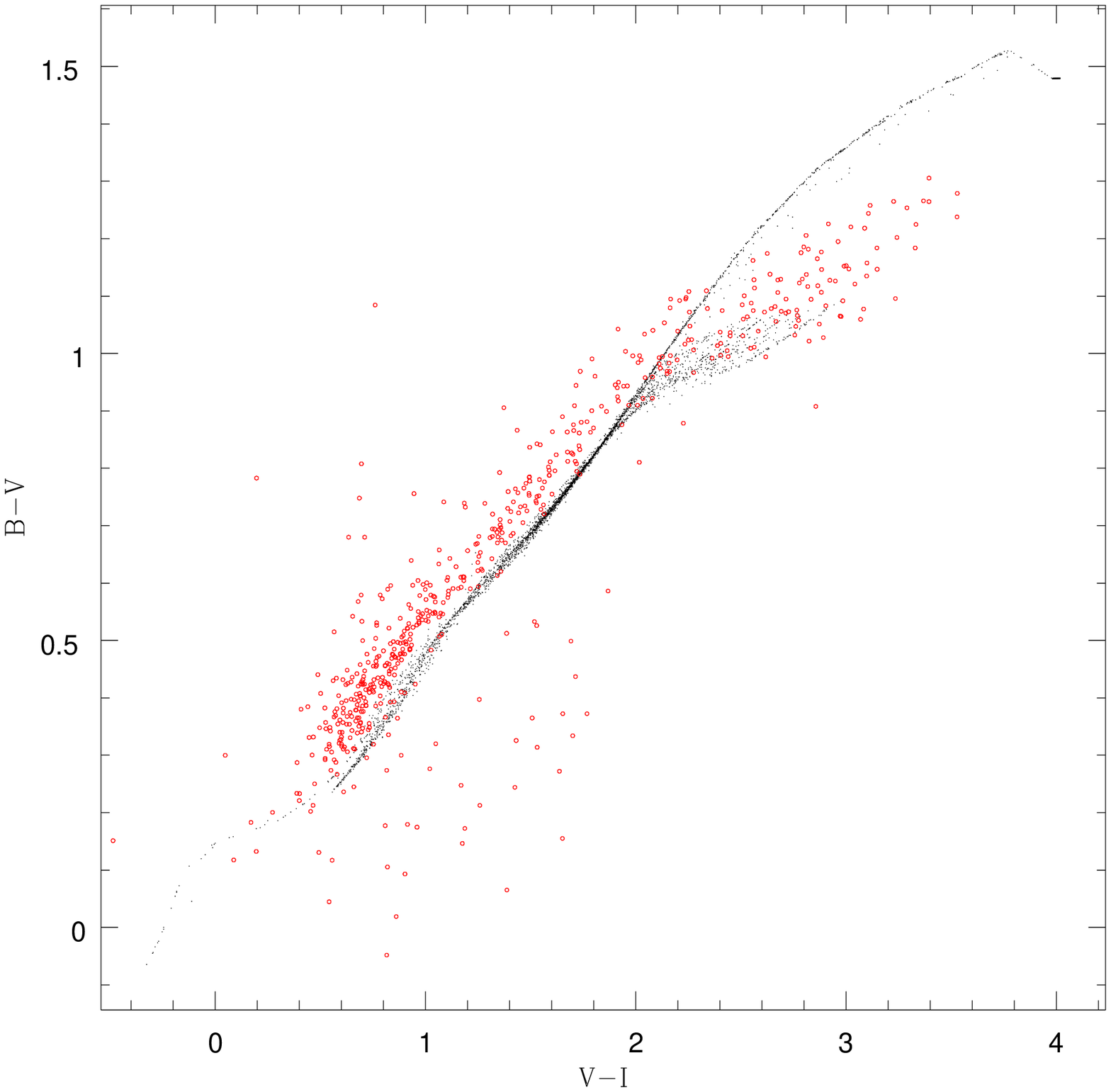}}
\resizebox{0.4\textwidth}{!}{\includegraphics{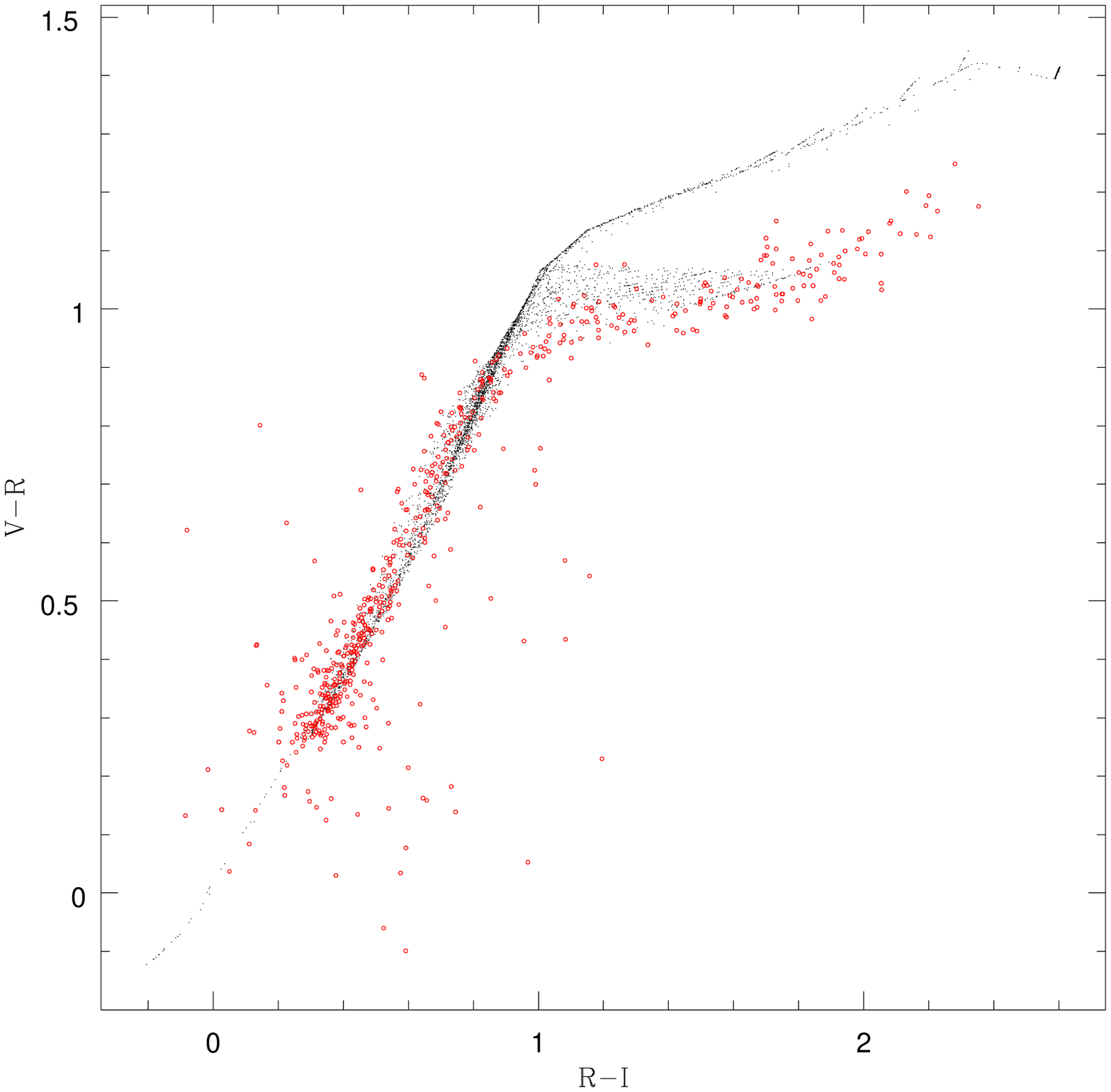}}
\caption{Color-color plot for stars (red points) selected in the field DEEP1b compared the model of \cite{Girardi2005} (black points).}
\label{fig:starcol1}
\end{figure*}

\begin{figure*}
\center
\resizebox{0.4\textwidth}{!}{\includegraphics{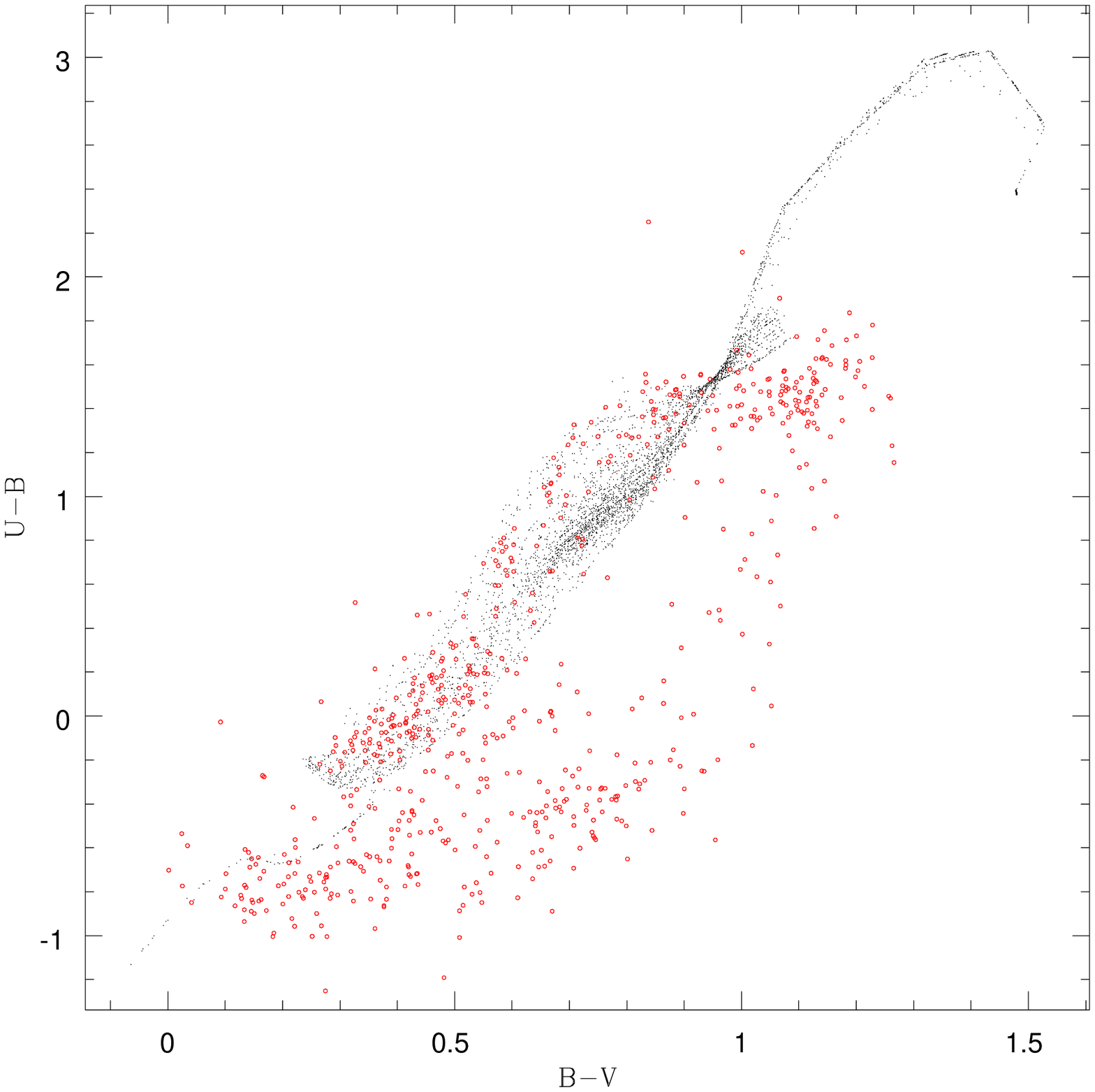}}
\resizebox{0.4\textwidth}{!}{\includegraphics{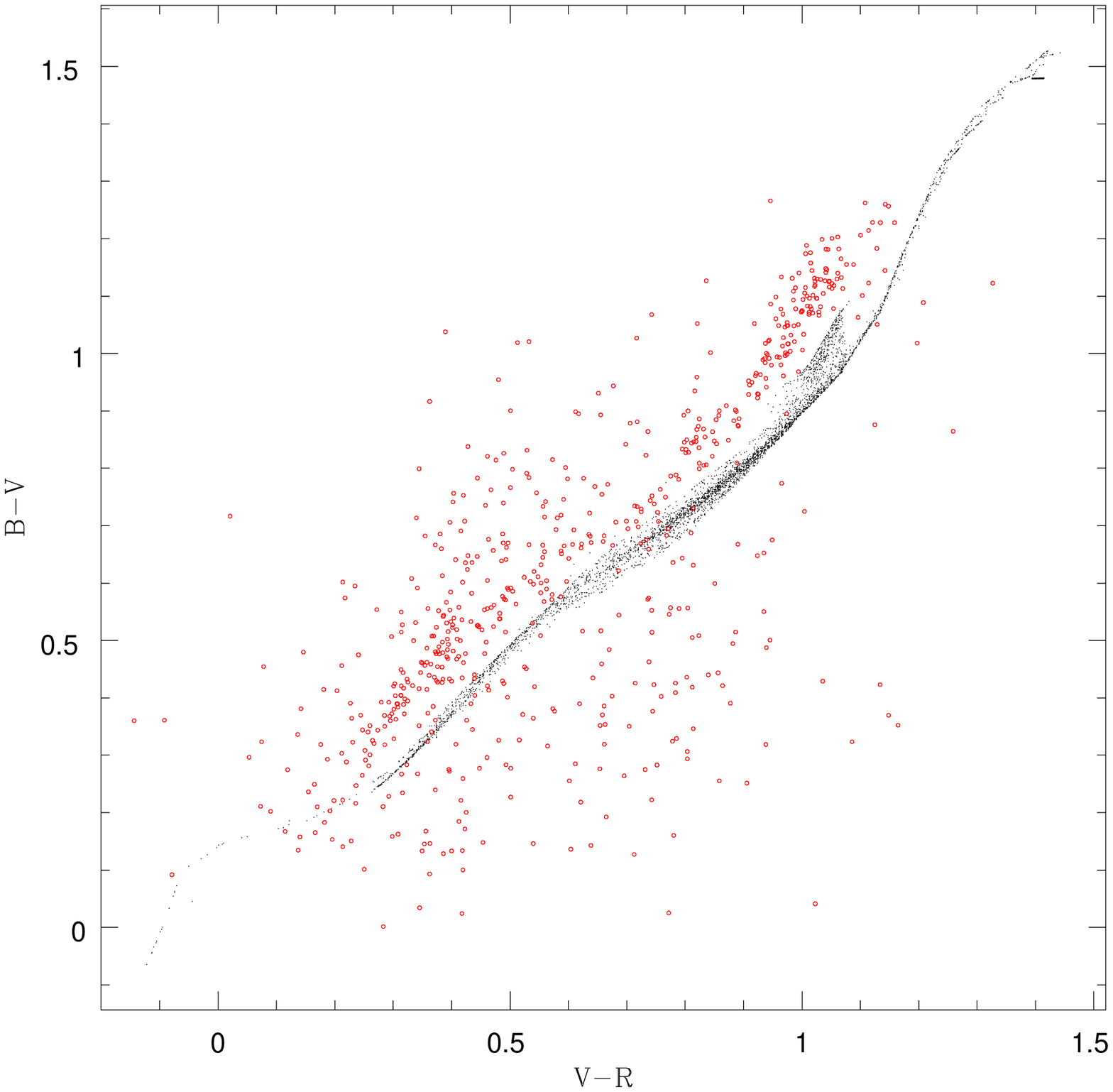}}

\resizebox{0.4\textwidth}{!}{\includegraphics{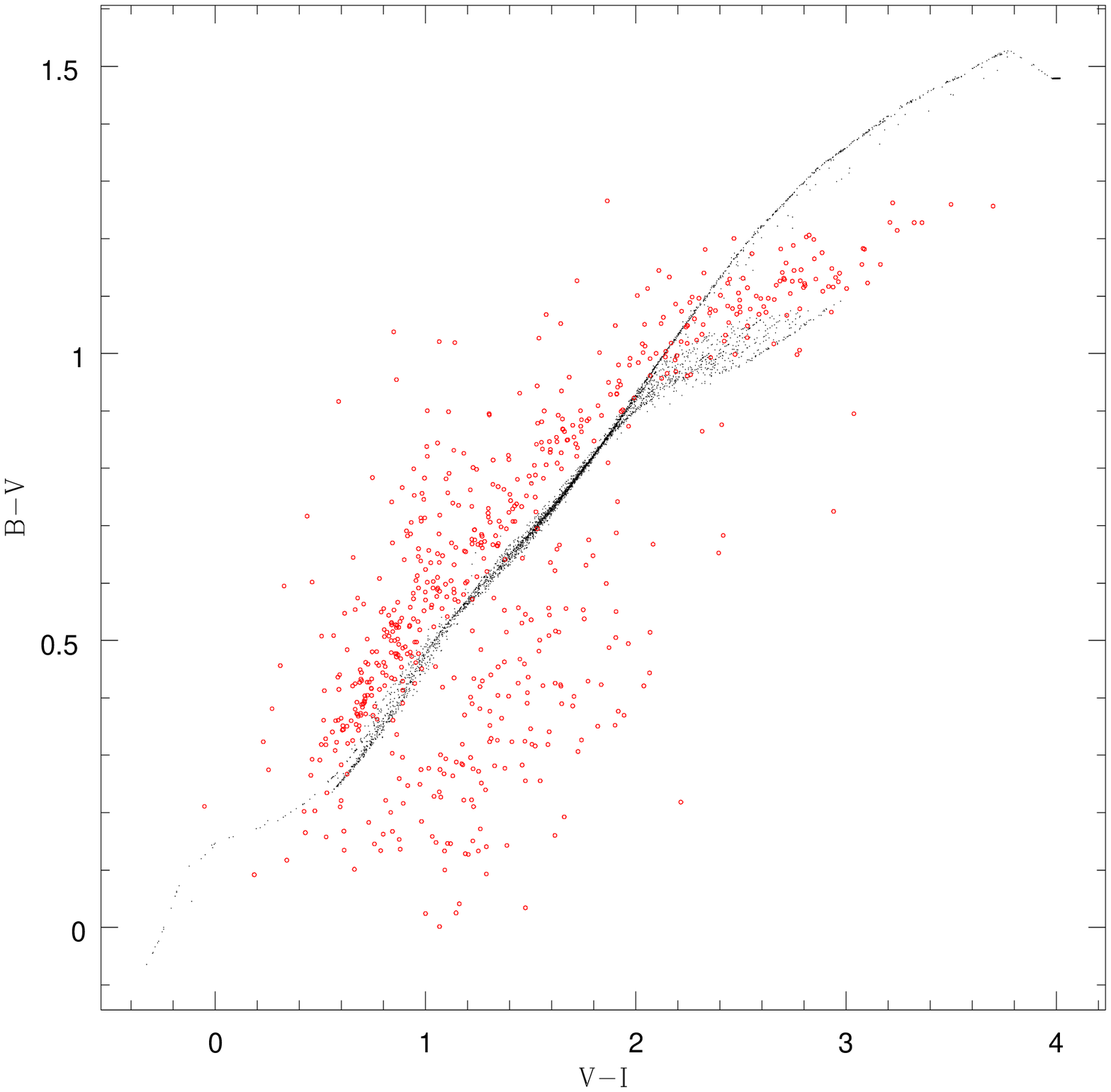}}
\resizebox{0.4\textwidth}{!}{\includegraphics{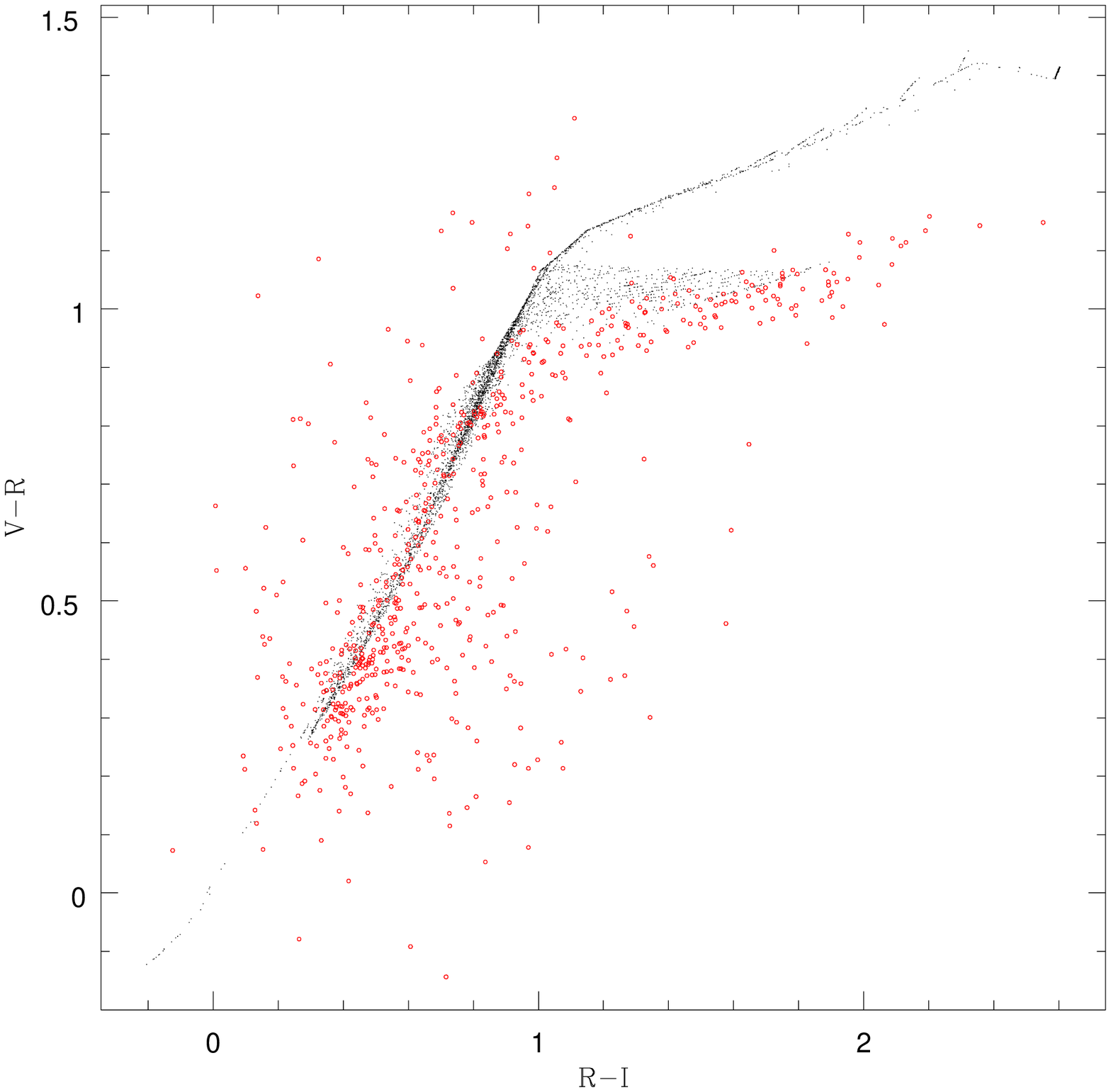}}
\caption{Same as Fig.~\ref{fig:starcol1} but for field DEEP2b.}
\label{fig:starcol2}
\end{figure*}

\begin{figure*}
\center
\resizebox{0.4\textwidth}{!}{\includegraphics{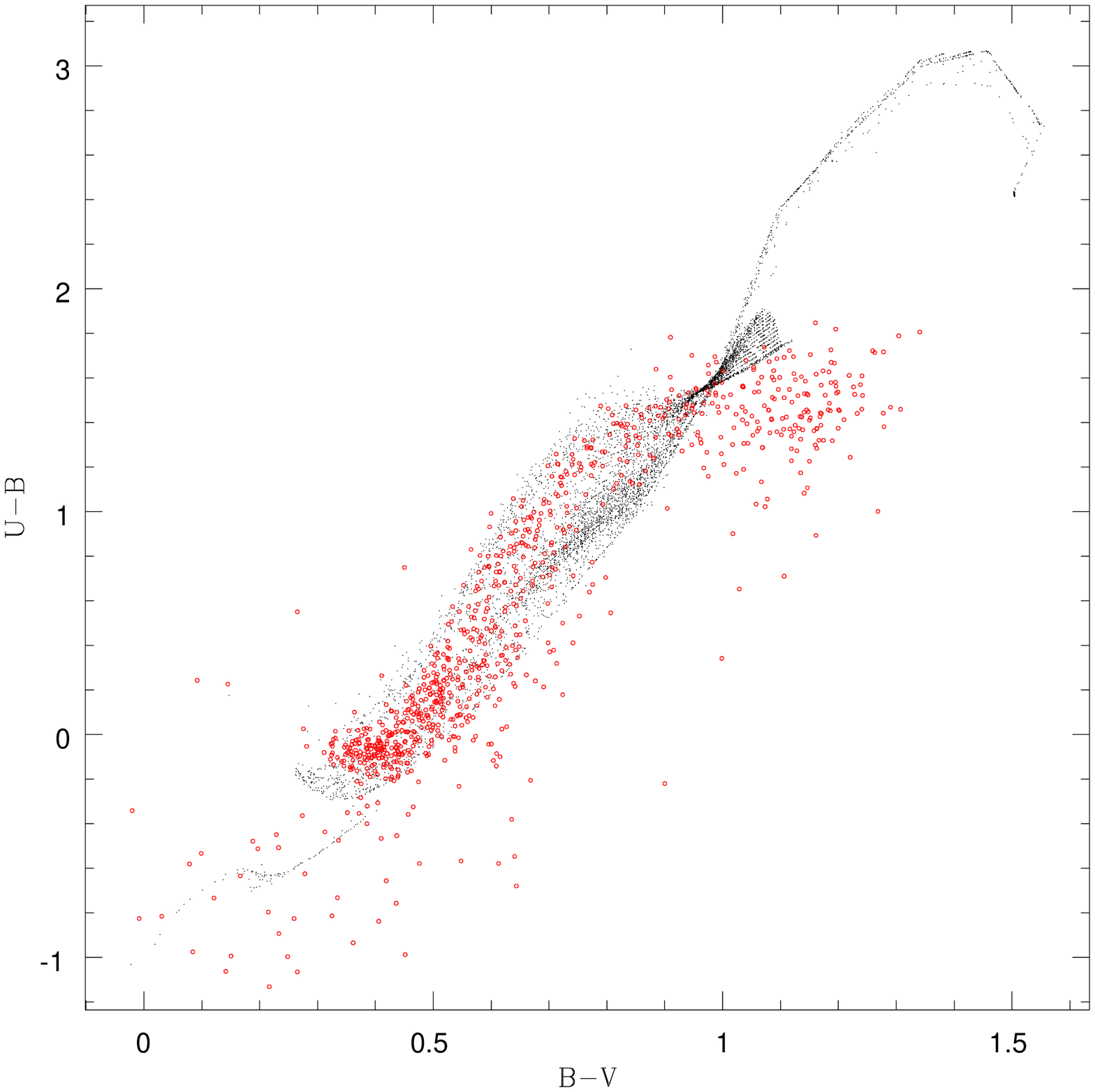}}
\resizebox{0.4\textwidth}{!}{\includegraphics{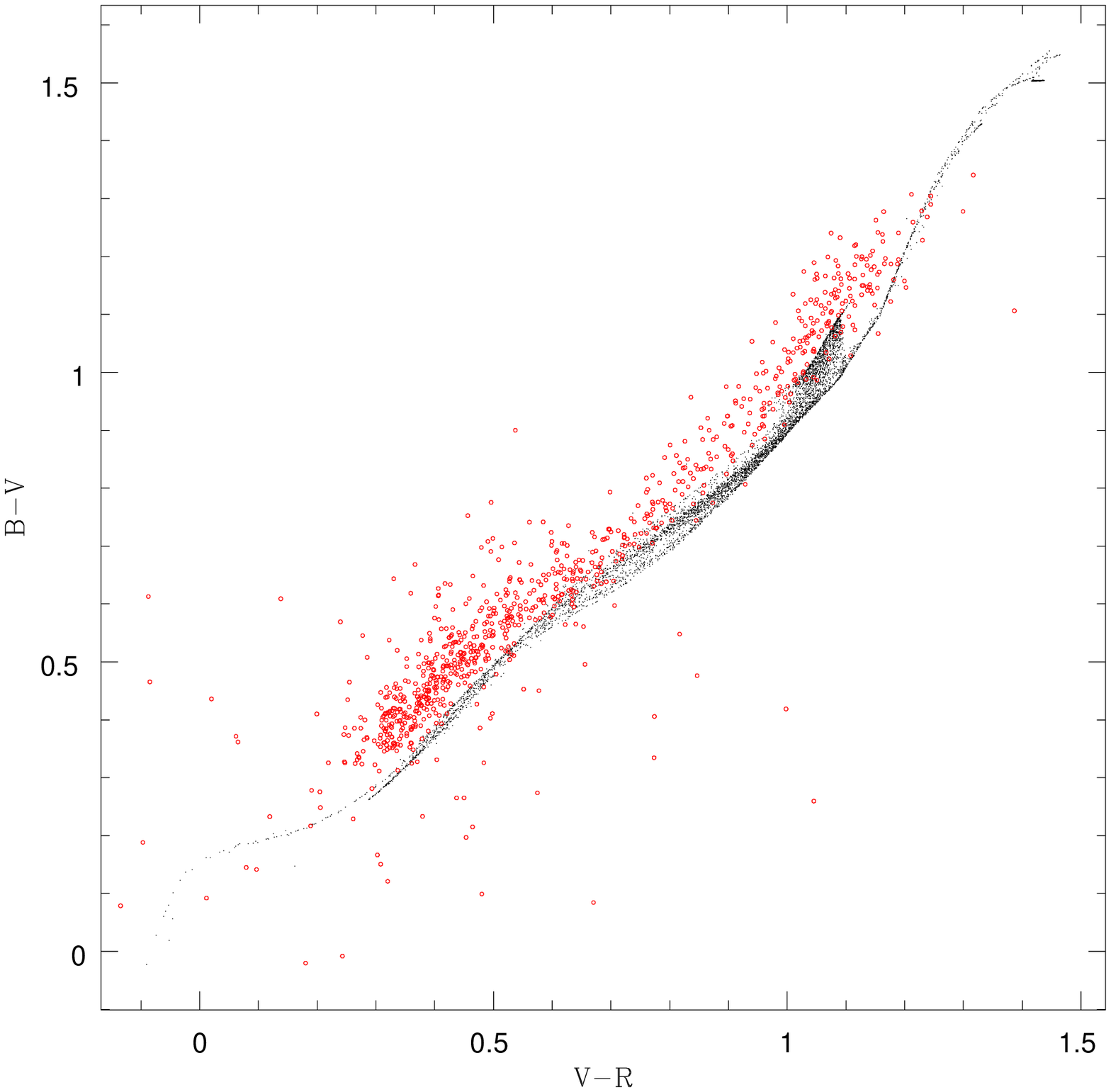}}

\resizebox{0.4\textwidth}{!}{\includegraphics{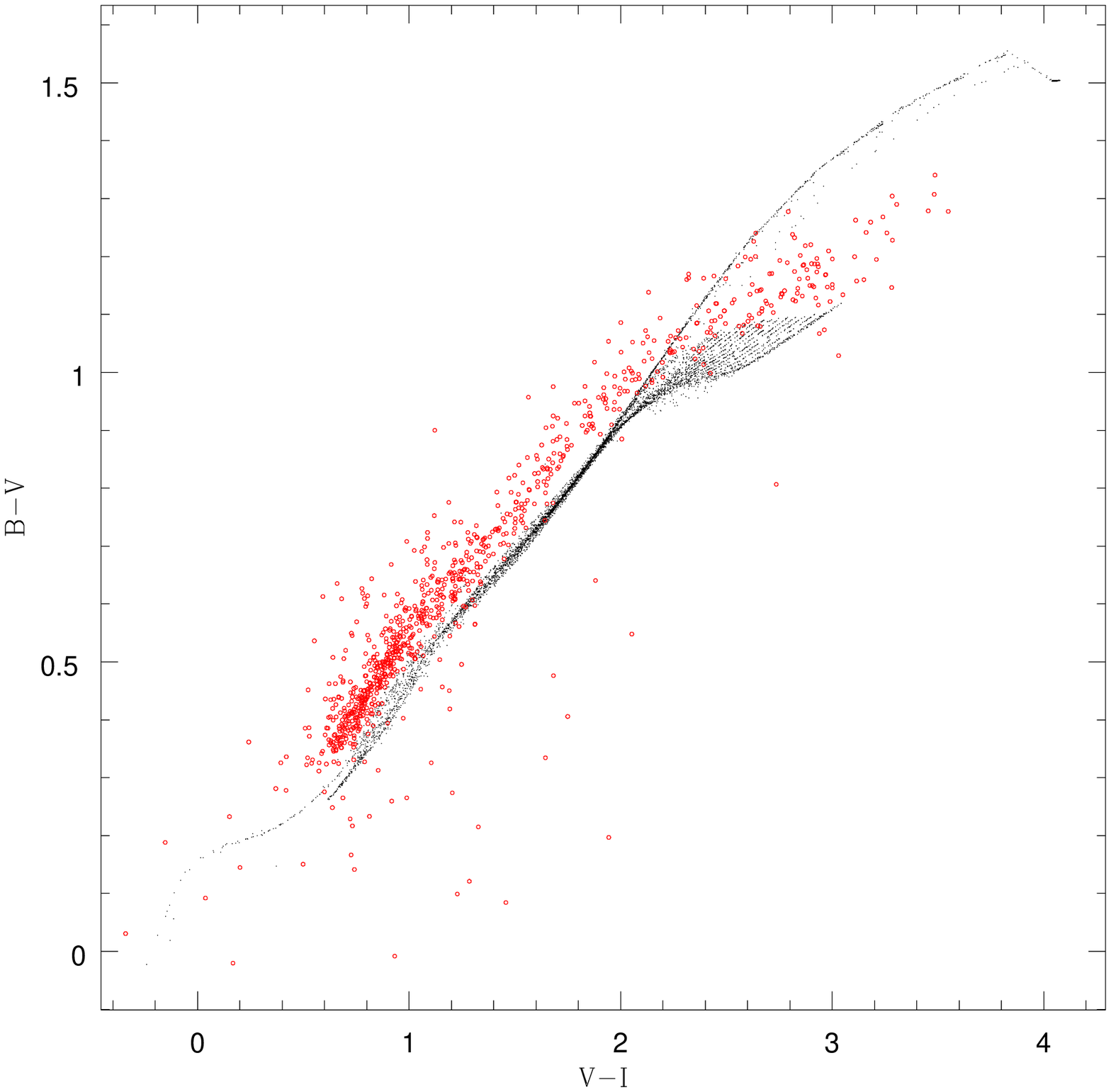}}
\resizebox{0.4\textwidth}{!}{\includegraphics{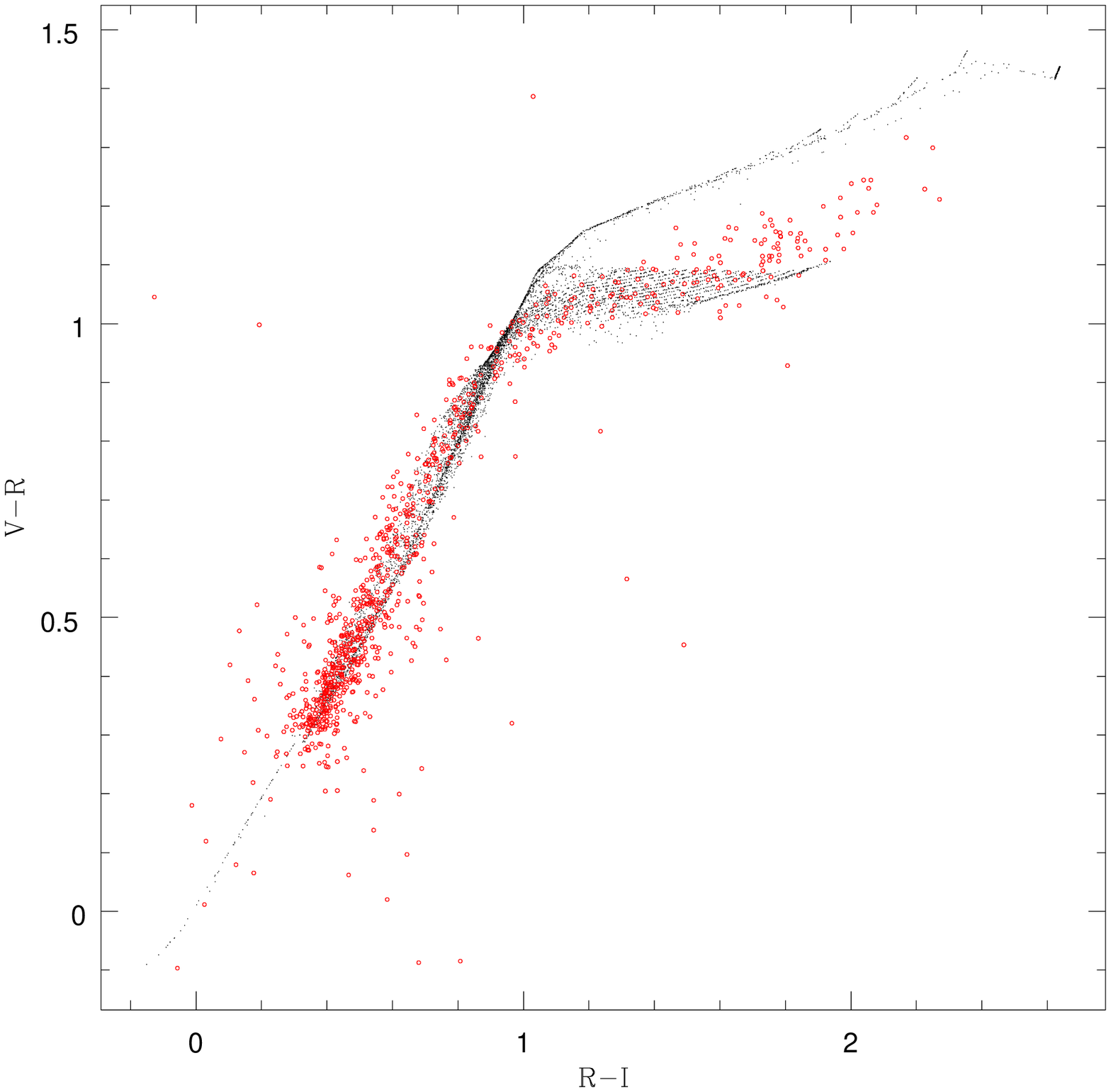}}
\caption{Same as Fig.~\ref{fig:starcol1} but for field DEEP3b.}
\label{fig:starcol3}
\end{figure*}

\begin{figure*}
\center
\resizebox{0.4\textwidth}{!}{\includegraphics{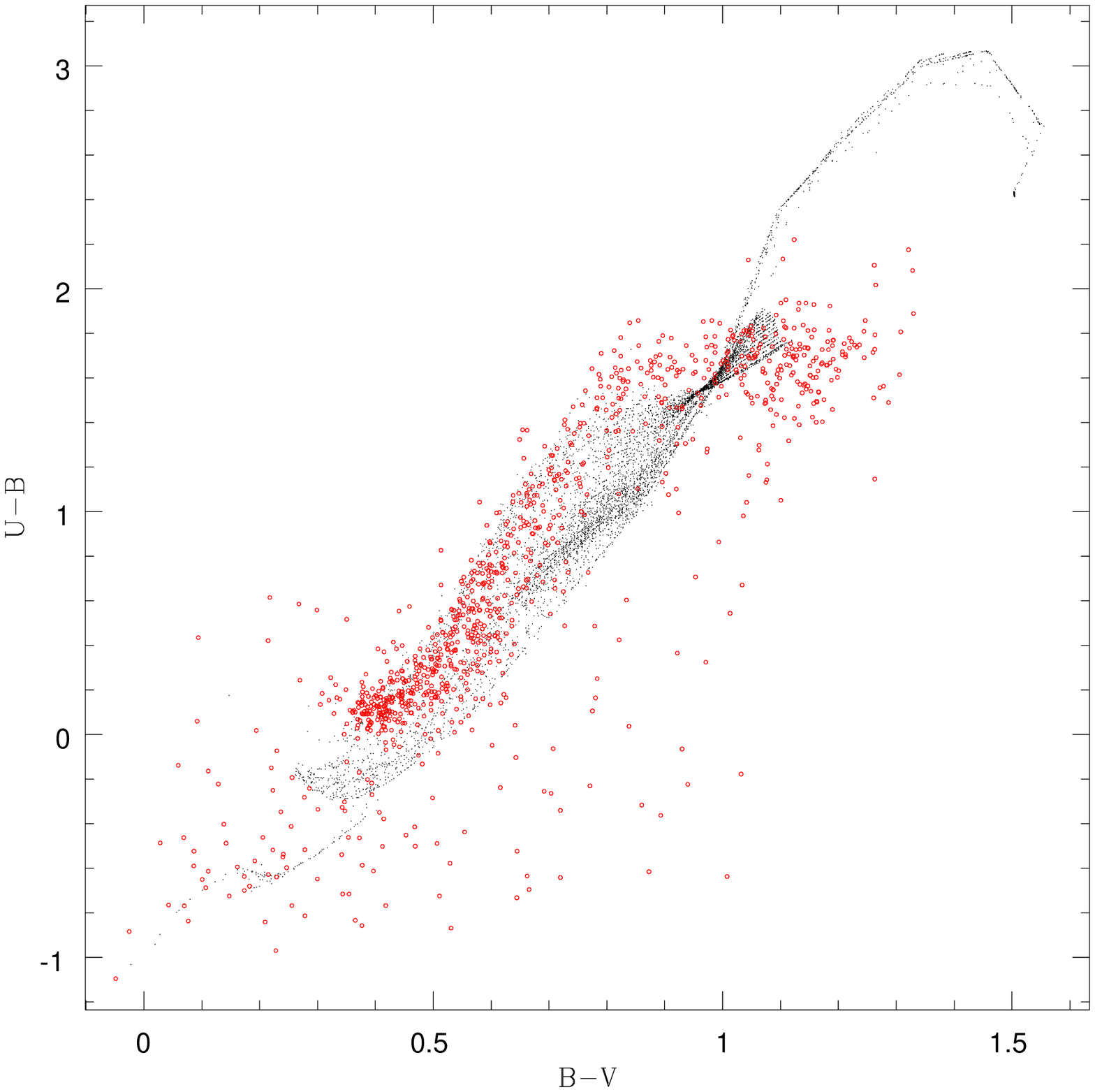}}
\resizebox{0.4\textwidth}{!}{\includegraphics{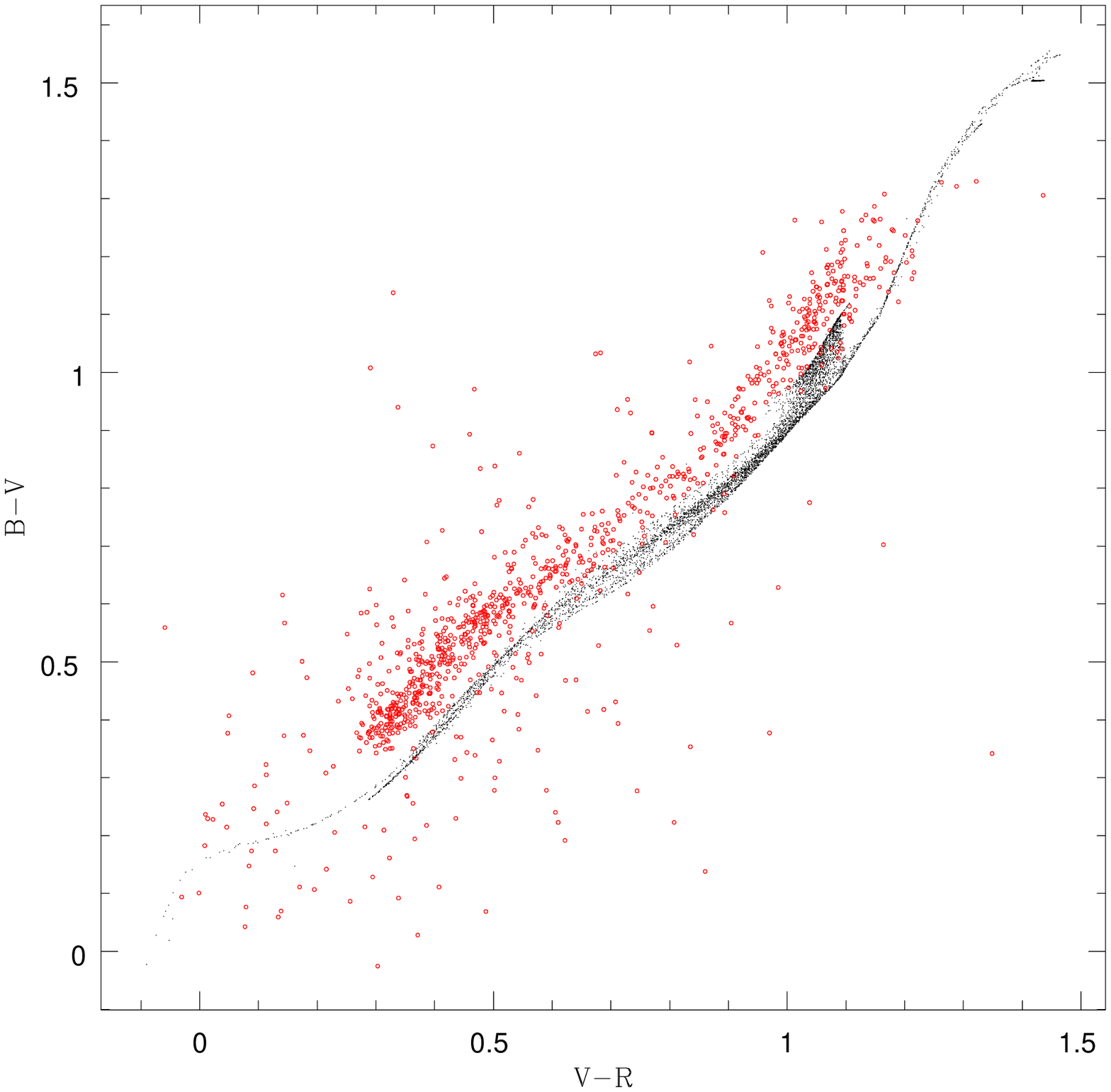}}

\resizebox{0.4\textwidth}{!}{\includegraphics{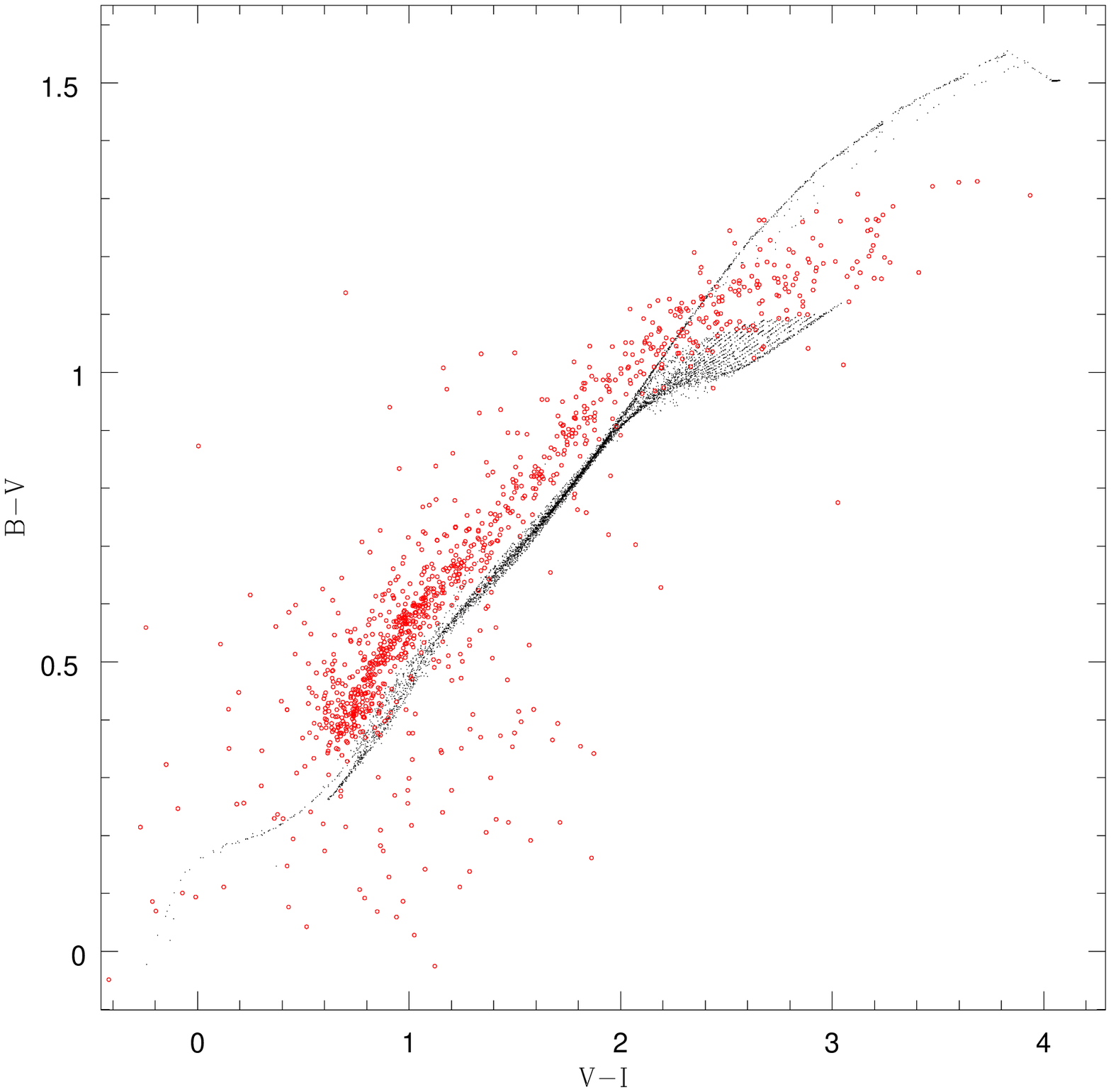}}
\resizebox{0.4\textwidth}{!}{\includegraphics{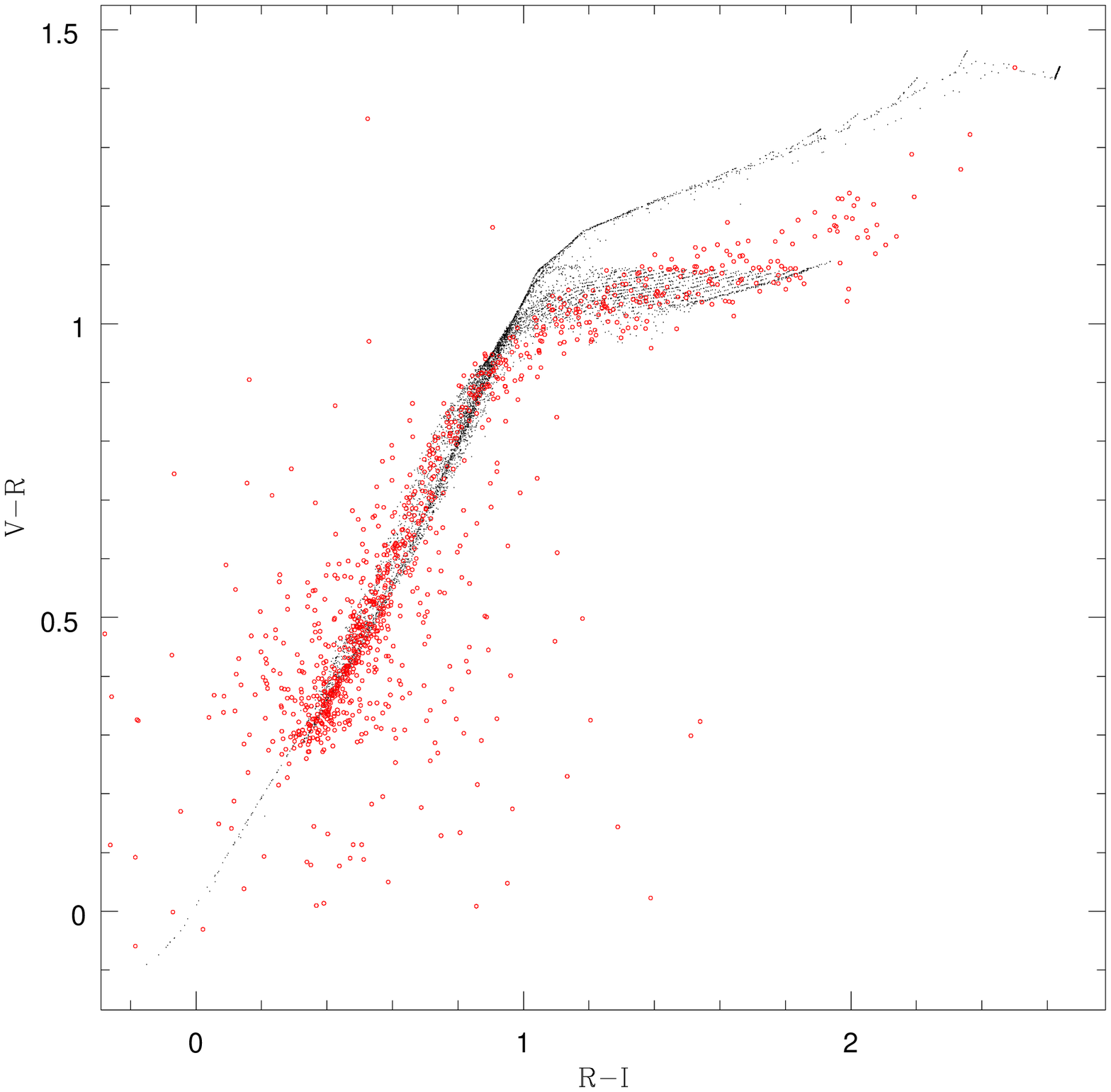}}
\caption{Same as Fig.~\ref{fig:starcol1} but for field DEEP3c.}
\label{fig:starcol4}
\end{figure*}

\bibliographystyle{aa}
\bibliography{DPS.am}
\end{document}